\documentclass[manuscript,screen,nonacm]{acmart}
\usepackage{subcaption}
\usepackage{booktabs}
\usepackage{multirow}
\usepackage{threeparttable}
\usepackage{siunitx}
\usepackage{rotating}
\usepackage{xcolor}

\usepackage{booktabs}
\usepackage{makecell}
\usepackage{amsmath}
\usepackage{url}
\usepackage{xcolor}
\usepackage{multirow}
\usepackage{threeparttable}
\usepackage[T1]{fontenc}
\usepackage{graphicx}
\usepackage{graphicx}
\usepackage{subcaption}
\usepackage{hyperref}

\AtBeginDocument{%
  }

\setcopyright{acmlicensed}
\copyrightyear{2018}
\acmYear{2018}
\acmDOI{XXXXXXX.XXXXXXX}
\acmJournal{JOCCH}
\acmISBN{978-1-4503-XXXX-X/2018/06}

\begin{document}

\title{Modeling Stylistic Co-evolution in Symbolic Music Heritage Collections}

\author{Yulong He}
\orcid{0000-0002-3001-0096}
\affiliation{%
\institution{St. Petersburg State University}
\streetaddress{University emb., 7/9}
\city{St. Petersburg}
\country{Russia}
}


\author{Ivan Smirnov}
\orcid{0009-0002-3690-4733}
\affiliation{%
\institution{ITMO University}
\streetaddress{Kronverkskiy av., 49}
\city{St. Petersburg}
\country{Russia}
}

\author{Yanming Li}
\orcid{0009-0008-4386-3101}
\affiliation{%
\institution{St. Petersburg State Institute of Culture}
\streetaddress{Palace emb., 2}
\city{St. Petersburg}
\country{Russia}
}


\renewcommand{\shortauthors}{He et al.}

\begin{abstract}
Digitized musical heritage collections offer new opportunities to examine how stylistic traditions change over historical time, but computational analyses often reduce musical works to static classifications or similarity scores. This article proposes a representation-to-dynamics framework for studying cross-cultural harmonic change in Western art music. Starting from symbolic chord sequences, we derive contextual chord embeddings, project work-level representations into a shared harmonic space, and reconstruct country-level trajectories through temporally causal Kalman filtering. These trajectories are then modeled with DeGroot and Friedkin--Johnsen dynamics, yielding interpretable influence-like networks and estimates of stylistic anchoring. We apply the framework to a curated corpus of 480 dated works from 1875 to 1940 across Russia, France, Germany, Austria, and a heterogeneous ''Others'' group. Models fitted on 1875--1925 are evaluated through recursive forecasts for 1930--1940, testing whether the estimated dependency structure remains informative across a potentially changing historical and stylistic regime. The estimated trajectories and influence patterns are broadly consistent with established music-historical accounts of late Romantic and early modernist exchange, including the close relationship between German and Austrian traditions and historically plausible cross-currents between Russian and French traditions. PCA-variance-weighted estimation provides modest improvements while preserving a single interpretable influence network. Rather than treating the estimated matrices as direct causal evidence, the framework offers a reproducible quantitative layer for cultural heritage research, complementing archival and musicological interpretation.
\end{abstract}



\maketitle

\section{Introduction}
\label{sec:introduction}

Musical style evolution offers a structured and quantifiable window into the mechanisms of cultural transmission. Across historical epochs, aesthetic traditions emerge through the interplay of endogenous creative trajectories and exogenous cross-border influences. Although musicologists have extensively documented these processes through qualitative scholarship, computational models that characterize stylistic diffusion across national traditions remain underdeveloped.

Current research exhibits a methodological divide. In Music Information Retrieval (MIR), advances in representation learning and style classification typically treat musical works as static objects or focus on univariate period discrimination~\cite{mauch2015computer,WeissMDM19_StyleEvolution_MusicaeScientiae}. Conversely, opinion dynamics models in computational social science provide rigorous frameworks for belief propagation and consensus formation~\cite{degroot1974reaching,Friedkin01011990}, but lack culturally grounded, high-dimensional observables. Bridging these domains requires (i) a validated proxy for national stylistic ``opinion,'' (ii) a robust signal extraction pipeline to handle sparse and noisy historical observations, and (iii) a dynamical model that explicitly disentangles endogenous tradition from exogenous influence.

\begin{figure*}[t]
\centering
\includegraphics[width=\textwidth]{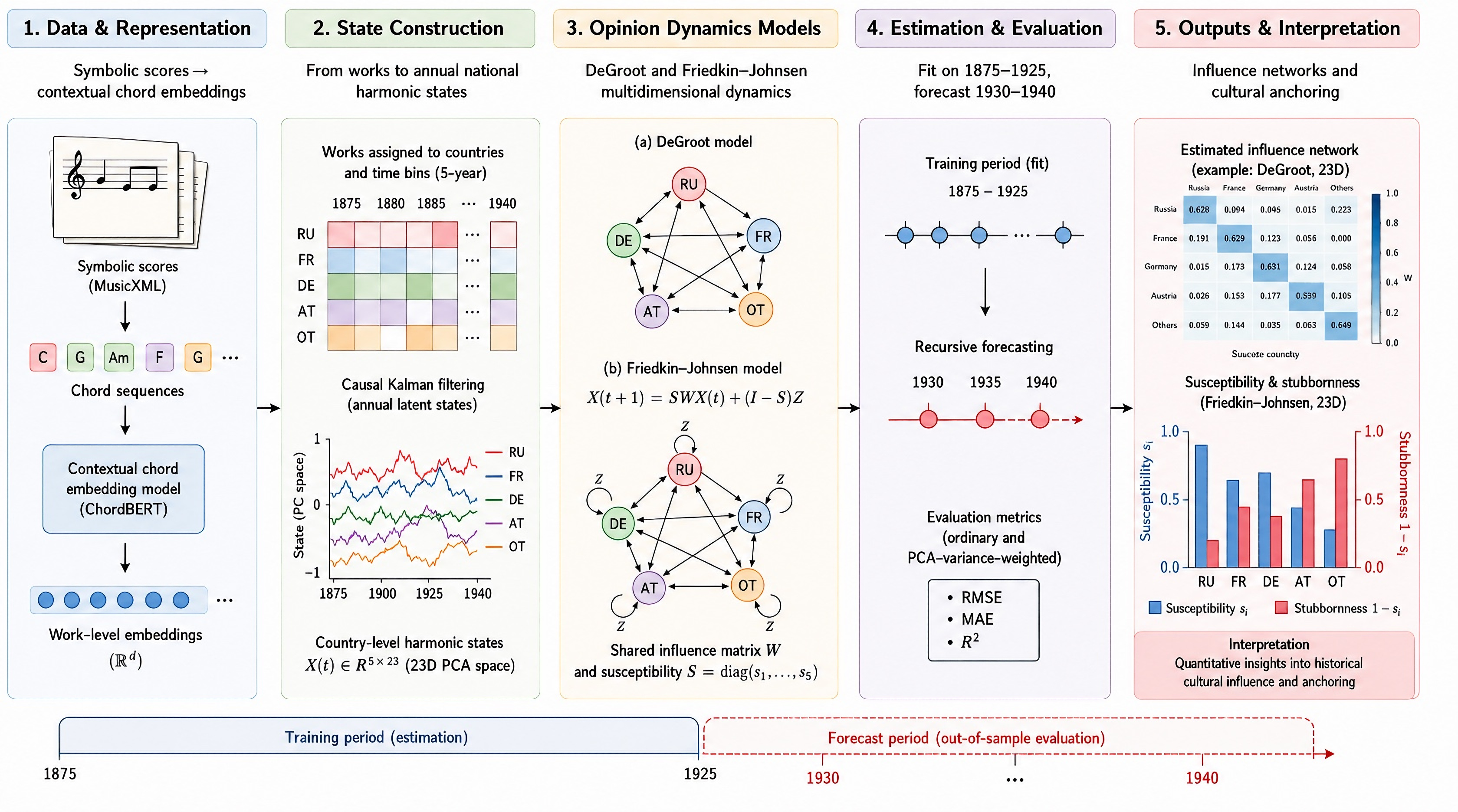}
\caption{Overview of the proposed framework.}
\label{fig:pipeline}
\end{figure*}

To address this gap, we present a unified computational framework (See Fig. \ref{fig:pipeline}) that integrates harmonic representation learning, signal processing, and opinion dynamics modeling. Using symbolic works from 1875–1940, a period spanning late Romanticism, musical nationalism, and early modernism, we construct a representation-to-dynamics framework for analyzing long-term stylistic change in digitized musical heritage collections. The dynamical models are fitted on 1875–1925 and evaluated retrospectively through recursive forecasts for 1930–1940.

Specifically, we (1) learn contextual chord embeddings to capture work-level harmonic syntax, (2) formulate cultural evolution as a latent signal recovery problem and apply filtering theory to denoise sparse temporal observations, and (3) formulate stylistic evolution as opinion dynamics, enabling interpretable modeling of cross-cultural harmonic dependencies and stylistic persistence. This enables quantitative estimation of asymmetric stylistic dependencies, including the close stylistic connection observed between German and Austrian traditions.

\noindent\textbf{Contributions.}
Our key contributions are as follows. First, we introduce a
representation-to-dynamics framework that links work-level harmonic embeddings, temporally filtered cultural trajectories, and interpretable opinion-dynamics models for studying long-term stylistic co-evolution in symbolic music heritage collections.

Second, we develop and evaluate leakage-free harmonic representations for symbolic musical works by training chord embedding models exclusively on external chord corpora, while keeping the historical target corpus fully separated from representation learning and model selection.

Third, we demonstrate the framework on a curated symbolic music corpus with source-backed work-level temporal annotations. The main analysis focuses on 480 works from 1875--1940 across five cultural groups (Russia, France, Germany, Austria, and Others), and examines how sparse observations, heterogeneous aggregation, and model assumptions affect the reconstructed stylistic trajectories.

Finally, we provide diagnostic extensions and open-source resources, a MIDI-based stress test, and released embeddings, signal extraction code, and dynamics estimation scripts to support replication and future research in computational cultural analysis.

The remainder of this paper is organized as follows: \S\ref{sec:related_work} reviews related work; \S\ref{sec:dataset}--\S\ref{sec:country_opinions} detail dataset construction, representation learning, and opinion dynamics analysis; and \S\ref{sec:discussion} discuss empirical findings, limitations, and broader implications. 
The appendices (\S\ref{app:pca-opinion-diagnostics}--
\S\ref{app:midi_bilstm}) provide supplementary diagnostics, including a MIDI-based stress test. The code is available at \url{https://github.com/hreyulog/music_opinion_dynamic}.
\section{Related Work}
\label{sec:related_work}

Our work intersects four active research areas: computational musicology, representation learning for symbolic music, opinion dynamics modeling, and cross-cultural influence analysis. We position our contributions at the intersection of these domains.

\subsection{Computational Musicology, Symbolic Representation, and Cultural Influence}

The quantitative analysis of musical style has a long tradition in musicology~\cite{meyer1956emotion,dahlhaus1989nineteenth}. Early computational approaches often relied on rule-based or hand-crafted musical features for tasks such as period classification and stylistic comparison~\cite{huron2006sweet}. More recent work has used large-scale symbolic and audio corpora to study stylistic change as a continuous historical process rather than as a set of discrete period labels~\cite{WeissMDM19_StyleEvolution_MusicaeScientiae,mauch2015computer}. For example, Mauch et al.~\cite{mauch2015computer} analyze harmonic and timbral evolution in the Billboard corpus from 1960 to 2010, identifying patterns of ``punctuated equilibria'' in popular music. Weiss et al.~\cite{WeissMDM19_StyleEvolution_MusicaeScientiae} extend this line of inquiry to classical music by using tonal complexity metrics to track stylistic trajectories across composer lifetimes.

In parallel, representation learning for symbolic music has become an important approach for modeling musical structure beyond manually designed features. Early approaches represented chord or note sequences with bag-of-words-style features~\cite{WeissM15_TonalComplexity_ICASSP}, while more recent methods adapt techniques from natural language processing, including Word2Vec~\cite{NIPS2013_9aa42b31} and Transformer architectures~\cite{vaswani2017attention}, to musical sequences. In the harmonic domain, Chord2Vec~\cite{10.1007/978-3-030-72914-1_12} learns distributed chord embeddings through Skip-gram training and demonstrates their usefulness for style classification and retrieval. These studies provide important foundations for large-scale musical style analysis, but most existing work either focuses on static representation, classification, or univariate temporal trends.

A related line of work in digital humanities and cultural analytics studies cultural influence across geographic, institutional, and social boundaries. Network-based approaches have been applied to literary translation flows~\cite{moretti2005graphs}, art auction prices~\cite{velthuis2005symbolic}, and scientific collaboration~\cite{doi:10.1073/pnas.98.2.404}. In music, prior work has mapped composer influence through biographical networks~\cite{doi:10.1126/science.1240064} or stylistic similarity~\cite{Serr__2012}. Such approaches are valuable for describing relationships among cultural actors, but they often infer influence post hoc from observed links or similarities, rather than modeling how stylistic trajectories evolve over time.

Our work builds on this literature by combining symbolic music representation learning with interpretable dynamical modeling. We compare a Skip-gram chord embedding baseline with Transformer-based chord representations and use the resulting work-level harmonic embeddings as observables for long-term stylistic trajectories. Instead of treating cross-cultural relationships as static similarities, we estimate influence-like dependency matrices that best explain the observed evolution of country-level harmonic states under opinion-dynamics models. This enables a cautious, model-based analysis of cross-cultural stylistic co-evolution in historical musical heritage data.

\subsection{Opinion Dynamics for Cultural Trajectories}
\label{subsec:opinion_dynamics_rw}

Opinion dynamics models provide a mathematical language for describing how individual or collective states evolve through interaction. 
Originating in sociology, social psychology, statistical physics and network science, these models are commonly used to study belief formation, consensus, polarization, susceptibility, and persistence in interconnected populations~\cite{degroot1974reaching, Friedkin01011990}. 
In the classical DeGroot model (DG model)~\cite{degroot1974reaching}, each agent updates its state (often referred to as an “opinion” in the social influence literature) by taking a weighted average of the current states of other agents. In our setting, the “agents” are cultural groups rather than individuals, and the “opinions” are not subjective beliefs but latent harmonic-style states derived from symbolic musical representations.
The influence structure is encoded by a row-stochastic matrix whose rows describe how strongly each agent depends on others. 
This linear formulation remains useful because it connects the observed temporal trajectories to an interpretable latent interaction structure.

The Friedkin-Johnsen (FJ) model extends dynamics by allowing each agent to remain partially attached to an intrinsic or initial opinion~\cite{Friedkin01011990}. 
Opinion change is therefore governed by both social influence and anchoring to an internal preference or prior state. 
The resulting susceptibility or stubbornness parameters are particularly relevant in settings where agents do not simply converge through repeated averaging, but instead retain persistent identities, traditions, or institutional constraints.

Although these models are often introduced for scalar opinions, later work has generalized opinion dynamics to vector-valued or multidimensional states, where each agent holds opinions across multiple dimensions~\cite{Parsegov_2017,scienceaag2624,Proskurnikov_2017}. 
This perspective is important for cultural and musical data, where stylistic variation cannot be adequately represented by a single scalar value. 
Related PCA-based approaches have constructed scalar opinion trajectories from a single principal component for the evolution of code repositories~\cite{HE2026102824}. 
In contrast, our framework retains multiple PCA coordinates and represents each cultural group by a vector-valued harmonic state. The resulting country-by-component matrix $\mathbf{X}(t)$ allows DG and FJ models to operate on multidimensional musical trajectories while still estimating a single interpretable cross-group dependency matrix.

Beyond their original use for human opinions, opinion-dynamics models have increasingly been adapted as interpretable tools for evolving traces of behavior, language, sentiment, and representation. 
Recent studies apply these models to real-world digital traces such as software repositories and social-media trajectories, where high-dimensional observations are first transformed into lower-dimensional temporal states and then modeled as interacting processes~\cite{HE2026102824,he2025opiniondynamicsmodelssentiment,he2026opiniondynamicsmutualinfluence}. 
These applications suggest that opinion dynamics can function as a general framework for modeling temporal dependence among latent states.

Applying opinion dynamics to historical music requires a change in the definition of the agent. 
In standard opinion-dynamics settings, agents are usually assumed to coexist within the same time window and to update their states through contemporaneous interaction. 
Individual composers do not satisfy this assumption in a long-horizon historical corpus: their lifetimes, active creative periods, and surviving works are unevenly distributed, and many composers never overlap temporally with one another. 
Treating composers as individual agents would, therefore, produce a highly sparse and discontinuous interaction system. For this reason, we define the agents at the level of national or cultural groups. These groups should not be interpreted as literal decision-making actors, but as analytical units that make it possible to aggregate works into continuous harmonic-style trajectories over multiple decades. 
This aggregation allows the DG and FJ models to be used in a restricted, descriptive sense: the estimated matrices summarize influence-like dependencies among group-level stylistic trajectories. This distinction is important for interpreting the model. The resulting dynamics do not claim that countries themselves form opinions or that cultural influence can be reduced to national interaction. 
Rather, country and cultural-group labels provide a practical temporal scaffold for studying how harmonic features associated with different traditions co-evolve over time. 
Contextual chord representations supply the musical evidence, Kalman filtering converts sparse work-level observations into temporally ordered group-level states, and opinion-dynamics models provide an interpretable summary of self-persistence, cross-group dependence, and anchoring in these long-term trajectories.

\section{Dataset}
\label{sec:dataset}

\subsection{Cross-Era Dataset}
\label{sec:cross_era_dataset}

We conduct our historical analysis on the Cross-Era dataset
\cite{MauchD10_DifficultChords_ISMIR,Weiss17_StyleAnalysis_PhD,WeissM15_TonalComplexity_ICASSP,WeissMD14_StyleClassification_ICMC,WeissMDM19_StyleEvolution_MusicaeScientiae,WeissS15_KeyDetectionStyle_ISMIR},
a collection of symbolic music annotations covering multiple
historical periods. The dataset provides chord sequences aligned with musical
works, along with metadata such as composer identity and nationality.

The Cross-Era dataset comprises 2,000 musical works by 70 composers. Each work
is represented as a sequence of chord labels obtained via automated chord
recognition at regular temporal intervals. These chord sequences form the basis
for constructing harmonic representations and country-level musical trajectories
in the subsequent analysis.


\begin{figure}[t]
    \centering
    \includegraphics[width=\textwidth]{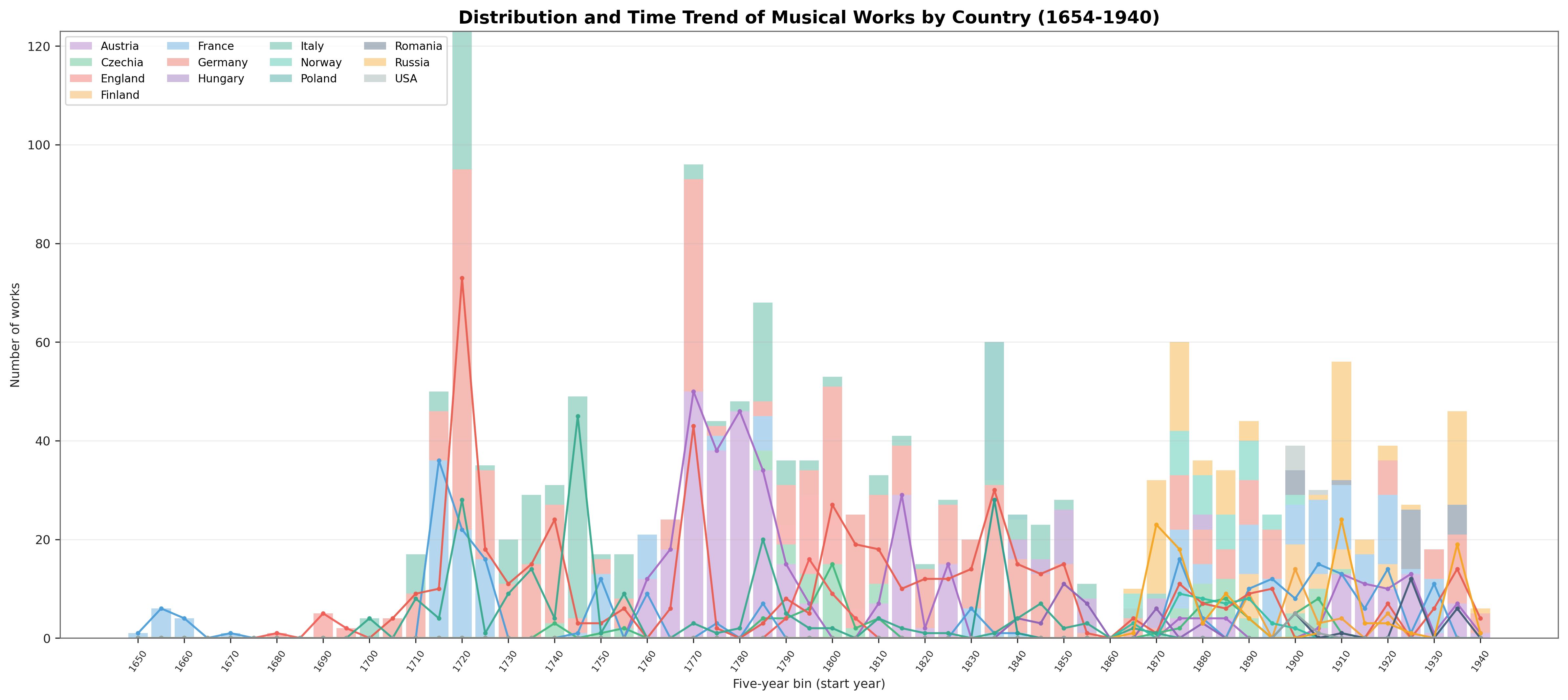}
    \caption{Distribution of musical works by country over time (1654--1940).}
    \label{fig:works_distribution}
\end{figure}

While composer-level metadata is readily available, work-level creation years
are often incomplete, inconsistent, or absent from the original metadata. To
enable fine-grained temporal analysis, we construct an auxiliary set of
work-level year annotations by collecting and cross-validating information from
multiple source-backed musicological resources. These sources include IMSLP work
pages and catalogues, Wikidata item properties, Wikipedia work pages and
composer work lists, RISM catalogue records, orchestra and concert-program
notes, publisher catalogues, and specialized music reference databases.

For each matched work, we record the inferred composition year or year range,
the supporting sources, a confidence label, and notes describing ambiguous
cases, source conflicts, or cases where only a publication, premiere, or
manuscript date is available. The resulting annotations are released as a
supplementary dataset on Hugging Face\footnote{Available at
\url{https://huggingface.co/datasets/StravynDynamics/cross-era-classical-work-years}}. This source-backed annotation process provides work-level temporal information for individual compositions, allowing us to conduct temporal analysis at the level of individual works rather than relying solely on aggregated composer-level timelines. Figure~\ref{fig:works_distribution} illustrates the distribution of works
across 13 countries over the extended temporal range. The stacked
bar chart shows the number of works per 5-year bin, while the line plot
highlights temporal trends for individual countries.

Based on this distribution, we restrict the subsequent opinion dynamics analysis
to the period 1875--1940. This choice is motivated by both data coverage and
music-historical considerations. First, our opinion dynamics model requires each
agent to be sufficiently represented across temporal windows, so that the
inferred trajectories are not dominated by sparse or missing observations. We
therefore select the four most consistently represented national
traditions---Germany, Austria, France, and Russia---as separate agents, while
grouping the remaining countries into a composite \textit{Others} agent.

Second, the period 1875--1940 corresponds to a major transitional phase from
late Romanticism to musical modernism. The years 1875--1900 are characterized
by late Romantic idioms and the flourishing of national schools, represented by
composers such as Brahms, Tchaikovsky, Dvo\v{r}\'{a}k, and the early Mahler.
The period 1900--1918 marks a phase of modernist breakthroughs, including
Impressionism, Expressionism, and the emergence of atonality, with figures such
as Debussy, Ravel, Schoenberg, and Stravinsky reshaping harmonic and formal
language. The interwar period, 1918--1940, further exhibits a pluralistic
modernist landscape in which neoclassicism, twelve-tone technique, nationalist
modernism, and jazz-inflected styles coexisted.

This time window is therefore particularly suitable for opinion dynamics
analysis: it captures a
historically dense transition in which multiple musical centers and aesthetic
movements competed, interacted, and influenced one another. Germany and Austria
trace a trajectory from late Romanticism toward Expressionism, atonality, and
twelve-tone organization; France moves from Impressionism toward neoclassicism;
Russia develops from nationalist traditions into modernist and \'{e}migr\'{e}
composer networks; and the \textit{Others} agent captures the heterogeneous
contributions of Czech, Hungarian, Italian, Nordic, American, and other national
or modernist styles.

\subsection{External Chord Corpora for Representation Learning}
\label{sec:external_chord_corpus}

The \textsc{Cross-Era} dataset is used as the target corpus for the
historical and cross-national analyses in this study. To avoid fitting the
harmonic representation directly to the target corpus, we train and select
the representation model using external chord corpora only. This design
allows the harmonic representations assigned to works in \textsc{Cross-Era}
to be obtained in a zero-shot, out-of-domain setting rather than being
optimized on the same data used for the downstream analysis.

We construct and evaluate a family of masked-language models for chord sequences, including a DeBERTa-based variant that we refer to as \textsc{ChordBERT}\footnote{Model available at
\url{https://huggingface.co/StravynDynamics/ChordBert}}. 
The goal of this stage is not to tune the representation on the downstream \textsc{Cross-Era} corpus, but to select a work-level harmonic representation through experiments on external chord datasets.

We use three external resources with distinct roles. 
First, \textsc{Chordonomicon}~\cite{kantarelis2024chordonomicondataset666000songs} is used as a large source-domain corpus for general chord representation learning. 
Although it consists primarily of contemporary and popular-music chord progressions, its large scale provides broad coverage of chord vocabulary and common progression patterns. 
Second, the \textsc{When-in-Rome} functional-harmony corpus~\cite{Gotham2023} is used as a classical-music domain-adaptation corpus. 
Because it contains symbolic functional-harmony annotations from the classical repertoire, it allows us to test whether continued masked-language pre-training on classical harmonic material improves the learned representations. 
Third, the \textsc{DCML} corpora are reserved exclusively for validation and model selection. 
They provide a clean, expert-annotated benchmark of classical harmonic analyses, enabling us to evaluate whether representations trained on different corpus combinations preserve musically meaningful harmonic structure.

Accordingly, we compare model variants trained under different data regimes, including source-domain training on \textsc{Chordonomicon}, classical-domain training on \textsc{When-in-Rome}, and combined source-plus-domain training. 
All variants are evaluated on the held-out \textsc{DCML} validation set, while \textsc{Cross-Era} is kept entirely separate from representation learning. 
This design ensures that the chord embedding model used for the downstream historical analysis is selected without using the target corpus for pre-training, domain adaptation, validation, model selection, or hyperparameter tuning.

\begin{table}[t]
\centering
\caption{Corpus statistics before and after filtering. Counts refer to progressions for Chordonomicon, works for When-in-Rome, scores for DCML, and tracks for Cross-Era.}
\label{tab:corpus_stats}
\begin{tabular}{llrr}
\hline
Corpus & Role & Before Items & Retained Items \\
\hline
Chordonomicon & Source pre-training & 679807 & 679807 \\
When-in-Rome & DAPT training & 1500 & 704 \\
DCML & Validation & 1283 & 367 \\
Cross-Era & Target analysis & 2000 & 2000 \\
\hline
\end{tabular}
\end{table}

To prevent leakage among the classical symbolic corpora, we construct mutually exclusive partitions with the following priority order: $
\textsc{Cross-Era} >
\textsc{DCML} >
\textsc{When-in-Rome}$. The \textsc{Cross-Era} dataset is held out entirely and all 2,000 works are retained for the final downstream analysis. 
The \textsc{DCML} corpora are used only for validation. 
Works overlapping with \textsc{Cross-Era} are removed from \textsc{DCML} using normalized composer names, catalogue identifiers such as BWV, K., Op., and WoO numbers, and fuzzy title matching. 
Ambiguous Mozart piano sonatas without reliable K\"ochel numbers are conservatively removed. 
This procedure yields a validation set of 367 works.

The classical domain-adaptation partition is constructed from the \textsc{When-in-Rome} functional-harmony corpus. 
Works overlapping with \textsc{Cross-Era} are removed. 
In addition, all composers represented in the final \textsc{DCML} validation set are held out from this training partition, preventing composer-level leakage between domain-adaptive training and validation. 
The resulting \textsc{When-in-Rome} training set contains 704 works.

Table~\ref{tab:corpus_stats} summarizes the corpus partitioning used in the representation-learning pipeline. 
\textsc{Chordonomicon} provides the large-scale contemporary/popular-music source domain, \textsc{When-in-Rome} provides the classical domain-adaptation partition, \textsc{DCML} is reserved for classical validation and model selection, and \textsc{Cross-Era} is retained in full as the held-out target corpus for the downstream historical analysis.

\section{Chord Representation Learning}
\label{sec:chord_representation}

\subsection{Chord Vocabulary Normalization}

All corpora are converted to the chord notation expected by the original \textsc{ChordBERT} tokenizer, a word-level whitespace tokenizer in which each chord and each special label (e.g., \texttt{<chorus\_3>}) are treated as a single token, resulting in a vocabulary of 3,230 tokens (obtained after training on the \textsc{Chordonomicon} dataset's note sequences). Representative conversions include \texttt{C\_maj} $\rightarrow$ \texttt{C}, \texttt{F\#\_min} $\rightarrow$ \texttt{Fsmin}, \texttt{A\_maj\_min7} $\rightarrow$ \texttt{A7}, and \texttt{C\_min\_min7} $\rightarrow$ \texttt{Cmin7}.

Some source chord labels are not covered by the tokenizer vocabulary. 
We rewrite these labels into supported chord-token forms using fixed conversion rules, while preserving the original labels in the metadata. 
No-chord events (conventionally denoted \textsc{N} \cite{inproceedings}) are retained in the raw records but are excluded during conversion, as they have no corresponding token in the \textsc{ChordBERT} vocabulary. We verified that all exported sequences consist exclusively of valid vocabulary tokens.
\subsection{Chord Representation Models}
\subsubsection{Word2Vec-style Skip-gram Chord Embeddings}

We additionally consider a Word2Vec-style Skip-gram model following the
Chord2Vec approach as an embedding baseline. Let a chord sequence be
denoted as $(c_1,c_2,\ldots,c_L)$,
where $L$ denotes the sequence length and $\ell$ indexes the position of
a centre chord. Let
$x_r^{\mathrm{pc}}(c)\in\{0,1\}$ indicate whether pitch class
$r\in\{1,\ldots,R\}$ is present in chord $c$, where $R$ denotes the
number of pitch classes.

Given a centre chord $c_\ell$ and a context chord
$c_{\ell+\Delta}$, the probability of pitch class $r$ occurring in the
context chord is modelled as

\[
p_r(c_\ell)
=
\sigma
\left(
\mathbf{a}_r^{\top}\mathbf{e}(c_\ell)+b_r
\right),
\]

where $\mathbf{e}(c_\ell)$ denotes the learned embedding of the centre
chord, $\mathbf{a}_r$ and $b_r$ are decoder parameters associated with
pitch class $r$, and $\sigma(\cdot)$ is the logistic sigmoid function.

Assuming that pitch classes are conditionally independent given the
centre chord, the Skip-gram objective is defined as

\begin{equation}
\mathcal{L}_{\mathrm{SG}}
=
-
\sum_{\ell=1}^{L}
\sum_{\Delta\in\mathcal{C}(\ell)}
\sum_{r=1}^{R}
\left[
x_r^{\mathrm{pc}}(c_{\ell+\Delta})
\log p_r(c_\ell)
+
(1-x_r^{\mathrm{pc}}(c_{\ell+\Delta}))
\log(1-p_r(c_\ell))
\right].
\end{equation}

where

\[
\mathcal{C}(\ell)
=
\{\Delta:
1\leq \ell+\Delta\leq L,\;
0<|\Delta|\leq u
\}
\]

denotes the set of valid context offsets within a context window of
radius $u$.

After training, a fixed-size representation of a musical work $m$ is
obtained by averaging the learned chord embeddings:

\begin{equation}
\mathbf{v}^{\mathrm{SG}}_m
=
\frac{1}{L_m}
\sum_{\ell=1}^{L_m}
\mathbf{e}(c_{m,\ell}),
\end{equation}

where $L_m$ denotes the number of chords in work $m$ and
$c_{m,\ell}$ denotes the $\ell$-th chord of that work.

\subsubsection{Transformer-based model}

We evaluate two Transformer-based masked-language models for chord
representation learning: a RoBERTa encoder \cite{RoBERTa} and a
DeBERTa encoder \cite{DeBERTa}. Both models use the same
\textsc{ChordBERT} tokenizer and vocabulary, so differences in
downstream performance can be attributed to the encoder architecture and
training regime rather than to differences in chord normalization. The initial \textsc{ChordBERT} checkpoint is trained on
\textsc{Chordonomicon}, which contains 679,807 contemporary chord
progressions. We use 543,845 progressions for training, 67,980 for
validation, and 67,982 for testing.

The models are trained with a masked language modelling (MLM) objective
\cite{devlin2019bertpretrainingdeepbidirectional}. Given a chord-token
sequence
$\mathbf{c}=(c_1,\ldots,c_L)$, and a set of masked positions $\mathcal{M}$, let $\tilde{\mathbf{c}}$ denote the corrupted input sequence after masking
the positions in $\mathcal{M}$. The MLM objective is

\begin{equation}
\mathcal{L}_{\mathrm{MLM}}
=
-\frac{1}{|\mathcal{M}|}
\sum_{\ell\in\mathcal{M}}
\log p_{\theta}(c_\ell|\tilde{\mathbf{c}}).
\end{equation}

This objective encourages the encoder to infer masked chord tokens from
their surrounding harmonic context.

In addition, we apply an unsupervised SimCSE-style contrastive objective
\cite{DBLP:journals/corr/abs-2104-08821}. For each chord sequence in a
batch, two dropout-augmented representations are generated, denoted as
$\mathbf{h}_b$ and $\mathbf{h}^{+}_b$. The contrastive objective is

\begin{equation}
\mathcal{L}_{\mathrm{CL}}
=
-\frac{1}{B}
\sum_{b=1}^{B}
\log
\frac{
\exp(
\mathrm{sim}(\mathbf{h}_b,\mathbf{h}^{+}_b)/\tau_c
)
}
{
\sum_{a=1}^{B}
\exp(
\mathrm{sim}(\mathbf{h}_b,\mathbf{h}^{+}_a)/\tau_c
)
},
\end{equation}

where $B$ denotes the batch size,
$\mathrm{sim}(\cdot,\cdot)$ is cosine similarity, and $\tau_c$ is the
temperature parameter.

The total training objective is

\begin{equation}
\mathcal{L}
=
\mathcal{L}_{\mathrm{MLM}}
+
\lambda\mathcal{L}_{\mathrm{CL}},
\end{equation}

where $\lambda=0.1$ controls the contribution of the contrastive
objective and is fixed for all experiments.

\subsection{Domain-Adaptive Pre-training}

We trained for 3 epochs using AdamW with a batch size of 64, an
initial learning rate of $5\times10^{-4}$, weight decay of 0.01, and
gradient clipping at 1.0. A cosine learning-rate schedule with a warm-up
phase covering the first 5\% of optimization steps is used. At each step,
15\% of tokens are dynamically selected for masking following the standard
80/10/10 corruption strategy: 80\% are replaced by \texttt{[MASK]}, 10\% by
a random chord token, and 10\% are left unchanged. 

\subsection{Evaluation Protocol and Model Selection}

\begin{table*}[t]
    \centering
    \caption{Chord- and work-level evaluation on the held-out DCML validation set.
Lower is better for MLM loss; higher is better for all other metrics.
``C'' denotes training on Chordonomicon, ``W'' denotes training on the
leakage-free When-in-Rome training partition, and ``C\&W'' denotes
training on both datasets. Dashes indicate inapplicable metrics:
Chord2Vec predicts pitch-class context rather than masked discrete
chord tokens and therefore does not support chord-level MLM evaluation.}
    \label{tab:chord_embedding_results}
    \scriptsize
    \setlength{\tabcolsep}{1.5pt}
    \renewcommand{\arraystretch}{0.92}
    \resizebox{\textwidth}{!}{%
    \begin{tabular}{@{}llcccccccccccc@{}}
        \toprule
        \multirow{3}{*}{\textbf{Level}}
        & \multirow{3}{*}{\textbf{Metric}}
        & \multicolumn{4}{c}{\textbf{Chord2Vec baseline}}
        & \multicolumn{8}{c}{\textbf{Transformer-based model}}\\
        && \multicolumn{4}{c}{\textbf{Skip-gram}}
        & \multicolumn{4}{c}{\textbf{RoBERTa}}
        & \multicolumn{4}{c}{\textbf{DeBERTa}} \\
        \cmidrule(lr){3-6}
        \cmidrule(lr){7-10}
        \cmidrule(lr){11-14}
        &
        & \textbf{Base}
        & \textbf{C}
        & \textbf{W}
        & \textbf{C\&W}
        & \textbf{Base}
        & \textbf{C}
        & \textbf{W}
        & \textbf{C\&W}
        & \textbf{Base}
        & \textbf{C}
        & \textbf{W}
        & \textbf{C\&W} \\
        \midrule

        \multirow{4}{*}{Chord}
        & MLM
        & -- & -- & -- & --
        & 8.1683 & 2.5042 & 6.5674 & 2.3951
        & 8.1270 & 2.0286 & 6.5660 & \textbf{1.7906} \\
        & Top-1
        & -- & -- & -- & --
        & 0.0003 & 0.3476 & 0.0555 & 0.3533
        & 0.0000 & 0.5326 & 0.1085 & \textbf{0.5472} \\
        & Top-10
        & -- & -- & -- & --
        & 0.0044 & 0.8367 & 0.3690 & 0.8418
        & 0.0003 & 0.8492 & 0.4485 & \textbf{0.8876} \\
        & MRR
        & -- & -- & -- & --
        & 0.0030 & 0.5075 & 0.1552 & 0.5173
        & 0.0014 & 0.6391 & 0.2051 & \textbf{0.6645} \\

        \midrule

        \multirow{4}{*}{\shortstack{Same\\work}}
        & R@1
        & 0.1825 & 0.1865 & 0.2024 & 0.1865
        & 0.2698 & 0.2619 & 0.2540 & 0.2738
        & 0.2302 & \textbf{0.3056} & 0.2381 & 0.2817 \\
        & R@10
        & 0.7540 & 0.7341 & 0.7381 & 0.7659
        & 0.8016 & 0.7976 & 0.8016 & 0.8056
        & 0.7937 & \textbf{0.8294} & 0.8056 & 0.8175 \\
        & MRR
        & 0.3619 & 0.3605 & 0.3606 & 0.3624
        & 0.4303 & 0.4340 & 0.4221 & 0.4441
        & 0.4089 & \textbf{0.4674} & 0.4128 & 0.4548 \\
        & mAP@10
        & 0.2049 & 0.1959 & 0.1875 & 0.2009
        & 0.2582 & 0.2483 & 0.2559 & 0.2503
        & 0.2544 & \textbf{0.2843} & 0.2494 & 0.2790 \\

        \midrule

        \multirow{3}{*}{\shortstack{Split\\half}}
        & R@1
        & 0.3270 & 0.2616 & 0.2779 & 0.2970
        & 0.5204 & 0.5368 & 0.5123 & \textbf{0.5395}
        & 0.4278 & 0.4959 & 0.4114 & 0.4768 \\
        & R@10
        & 0.7275 & 0.6975 & 0.6812 & 0.7193
        & 0.8692 & 0.9237 & 0.8692 & \textbf{0.9292}
        & 0.8011 & 0.8556 & 0.7956 & 0.8365 \\
        & MRR
        & 0.4541 & 0.3956 & 0.4096 & 0.4211
        & 0.6295 & 0.6647 & 0.6215 & \textbf{0.6686}
        & 0.5499 & 0.6082 & 0.5372 & 0.5940 \\

        \bottomrule
    \end{tabular}%
    }
\end{table*}

We evaluate three families of chord representation models: a non-contextual Skip-gram baseline, a RoBERTa-based masked-language model, and a DeBERTa-based masked-language model that we refer to as \textsc{ChordBERT}. 
The Transformer-based models are evaluated at both the chord level and the work level, whereas the Skip-gram baseline is evaluated only at the work level because it does not define a masked-token prediction distribution over the full chord vocabulary.
For chord-level evaluation, we apply dynamic masking to the held-out \textsc{DCML} validation set. The evaluation comprises 11,552 masked chord-token instances and is conducted for the RoBERTa- and DeBERTa-based models. We report masked-language-modeling (MLM) loss, Top-1 accuracy, Top-10 accuracy, and mean reciprocal rank (MRR). Because each masked position has a single ground-truth token, mean average precision is not reported at the chord level, as it is equivalent to MRR in this setting.
For work-level evaluation, each model is converted into a fixed-size representation. For the Transformer-based models, token-level hidden states are mean-pooled within each input chunk after excluding special and padding tokens. The resulting chunk representations are then aggregated using the number of valid chord tokens in each chunk as weights. For the Skip-gram baseline, the work-level representation is obtained by averaging the learned chord embeddings over the complete chord sequence.
We evaluate these representations using same-work movement retrieval and split-half identity retrieval. For both protocols, we report Recall@1, Recall@10, and MRR. We additionally report mAP@10 for same-work movement retrieval, where a query may have multiple relevant movements from the same composition. Formal definitions of all chord-level and work-level evaluation metrics are provided in Appendix~\ref{app:representation-metrics}.

To quantify the effect of pre-training, we also compare against randomly initialized Transformer baselines with the same tokenizer, architecture, pooling operation, and validation data as their corresponding trained models. Table~\ref{tab:chord_embedding_results} reports the chord- and work-level evaluation results. 
Continued masked-language pre-training on the \textsc{When-in-Rome} corpus improves chord-level prediction for the Transformer models. 
For the DeBERTa-based \textsc{ChordBERT} model, adding \textsc{When-in-Rome} training after \textsc{Chordonomicon} reduces MLM loss from 2.0286 to 1.7906, increases Top-$1$ accuracy from 0.5326 to 0.5472, increases Top-$10$ accuracy from 0.8492 to 0.8876, and improves MRR from 0.6391 to 0.6645. 
A similar pattern is observed for the RoBERTa-based model, indicating that classical-domain adaptation improves local masked-chord prediction.

The work-level retrieval results show a different pattern. 
The DeBERTa-based \textsc{ChordBERT} model trained on \textsc{Chordonomicon} achieves the strongest same-work movement retrieval performance among the evaluated representations, with Recall@1 of 0.3056, Recall@10 of 0.8294, MRR of 0.4674, and mAP@10 of 0.2843. 
These metrics are most directly aligned with our downstream use case, where each musical work must be represented by a stable work-level harmonic embedding. 
Although continued \textsc{When-in-Rome} pre-training improves chord-level prediction, it slightly lowers the DeBERTa work-level retrieval scores. 
This suggests that the masked-language objective improves local harmonic prediction but does not necessarily optimize the mean-pooled work-level representation used in the historical dynamics analysis.

For the downstream analysis, we therefore select the DeBERTa-based \textsc{ChordBERT} variant trained on \textsc{Chordonomicon} as the main harmonic embedding model. 
The RoBERTa-based models provide competitive split-half retrieval performance, and the Skip-gram model provides a non-contextual baseline, but the \textsc{Chordonomicon}-trained \textsc{ChordBERT} model offers the best same-work organization of the embedding space. 
We use its work-level embeddings as the harmonic representation layer for the subsequent opinion-dynamics analysis.

\section{Opinion-Dynamics Analysis}
\label{sec:country_opinions}

Based on the temporal distribution of annotated works
(Fig.~\ref{fig:works_distribution}), we retain Russia, France, Germany and Austria as separate cultural groups because they provide the most continuous coverage across the 1875--1940 analysis window. Works from the remaining represented countries---Norway, Finland, Czechia, Italy, Hungary, and the United States---are pooled into an ``Others'' category to avoid discarding sparsely represented material. This category is used only as a pooled comparison group and is not interpreted as a coherent national or cultural tradition.

Annual observations are sparse for several groups. We therefore sample the reconstructed trajectories at five-year intervals, which provides a compromise between temporal resolution and the number of observed works
available at each time point. The resulting analysis grid spans 1875--1940. We use the states from 1875--1925 for parameter estimation and reserve the final three states, corresponding to 1930, 1935, and 1940, as a chronological holdout set for recursive retrospective
evaluation.

\subsection{Country-level opinion state construction}
\label{sec:country_opinions}

We construct country-level opinion states from work-level harmonic representations. 
In this paper, an ``opinion'' does not denote a subjective attitude, but a multidimensional harmonic-style state derived from symbolic chord sequences. 
Let $K$ denote the number of cultural groups and let $D$ denote the number of retained PCA dimensions.

The construction proceeds in four steps. 
First, each work is encoded as a contextual chord embedding. 
Second, work embeddings are projected into a low-dimensional PCA space. 
Third, each PCA coordinate is normalized to a common scale. 
Fourth, annual country-level observations are smoothed through a causal Kalman filter. 
The resulting filtered states form the matrix $\mathbf{X}(t)$ used by the DG and FJ models.

\subsubsection{Step 1: Work embedding with \textsc{ChordBERT}}

Each musical work is represented by a 256-dimensional contextual embedding
obtained from \textsc{ChordBERT}, a DeBERTa-based masked-language model for
chord sequences. The chord sequence of work $w$ is tokenized and, because the
model has a maximum input length of 256 tokens, truncated to at most 254 chord
tokens, leaving space for the tokenizer-added boundary tokens.

Let $L_{w}$ denote the tokenized length of work $w$ and let
$m_{w\ell}\in\{0,1\}$ be the attention-mask value at token position $\ell$. The
work-level embedding is obtained by masked mean pooling over non-padding
positions:
\begin{equation}
    \mathbf{v}_{w}
    =
    \frac{
        \sum_{\ell=1}^{L_{w}} m_{w\ell}\,\mathbf{h}_{w\ell}
    }{
        \sum_{\ell=1}^{L_{w}} m_{w\ell}
    },
    \label{eq:work_embedding}
\end{equation}
where $\mathbf{h}_{w\ell}\in\mathbb{R}^{256}$ is the last-layer hidden state of
the DeBERTa encoder at position $\ell$ and padding positions are excluded by the
attention mask. This produces one fixed-size contextual harmonic embedding
$\mathbf{v}_{w}\in\mathbb{R}^{256}$ for each work.

\subsubsection{Step 2: PCA dimensionality reduction}

Principal component analysis is fitted on the \textsc{ChordBERT} embeddings of works dated between 1875 and 1940. 
Let
\[
    \mathcal{W}_{1875:1940}
    =
    \{w: 1875 \leq \tau_w \leq 1940\}
\]
denote this PCA fitting sample, where $\tau_w$ is the annotated composition year of work $w$. 
Let $\mathbf{V}_{\mathrm{fit}}\in\mathbb{R}^{480\times256}$ be the corresponding matrix of work-level embeddings. 
PCA maps each work embedding to an orthogonal coordinate vector:
\begin{equation}
    \mathbf{z}_{w}
    =
    (z_{w1},z_{w2},\ldots,z_{wD})
    =
    (\mathbf{v}_{w}-\bar{\mathbf{v}})\mathbf{A}_{D},
    \label{eq:pca_projection}
\end{equation}
where $\bar{\mathbf{v}}$ is the mean embedding in the PCA fitting sample and $\mathbf{A}_{D}\in\mathbb{R}^{256\times D}$ contains the first $D$ principal directions.

\begin{figure*}[t]
\centering
\includegraphics[scale=0.4]{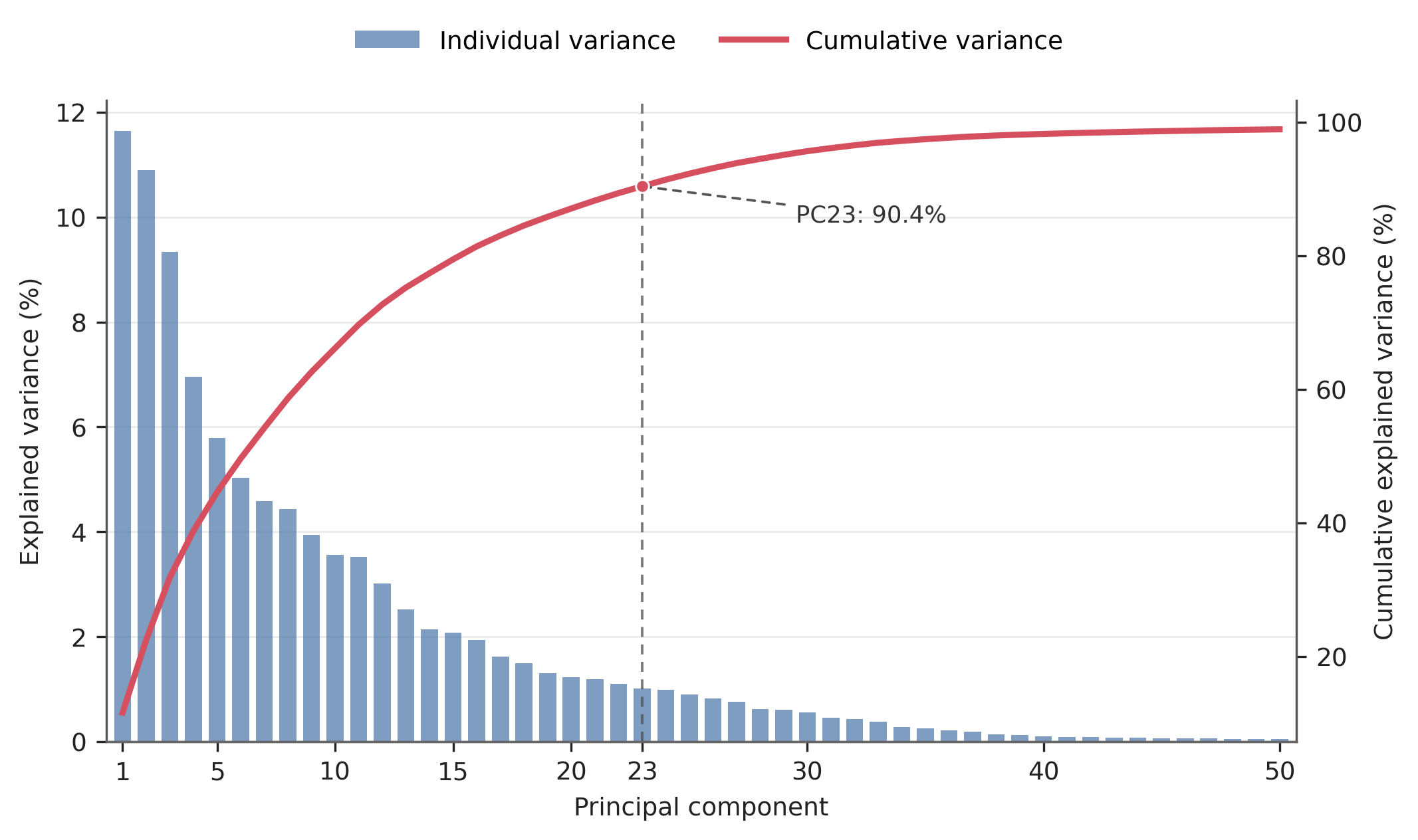}
\caption{Explained variance of the retained PCA components.}
\label{PCA_performance}
\end{figure*}

We retain the first $D=23$ components. 
Let $\rho_d$ denote the explained-variance ratio of component $d$. 
The retained components satisfy $\sum_{d=1}^{D}\rho_d = 0.903726$ and therefore explain approximately $90.37\%$ of the total embedding variance, as shown in Fig.~\ref{PCA_performance}. The PCA transformation fitted on the 1875--1940 sample is also applied to works preceding 1875. 
These earlier observations are used only to initialize the subsequent forward Kalman-filter recursion.

\subsubsection{Step 3: Component-wise min--max normalization}

After PCA projection, each retained component is normalized independently. 
For component $d$, the minimum and maximum are computed over the PCA fitting sample:
\begin{equation}
    z_{d}^{\min}
    =
    \min_{w\in\mathcal{W}_{1875:1940}} z_{wd},
    \qquad
    z_{d}^{\max}
    =
    \max_{w\in\mathcal{W}_{1875:1940}} z_{wd}.
    \label{eq:pca_minmax_values}
\end{equation}

The normalized score of work $w$ on component $d$ is
\begin{equation}
    \tilde{z}_{wd}
    =
    \frac{
        z_{wd}-z_{d}^{\min}
    }{
        z_{d}^{\max}-z_{d}^{\min}
    },
    \qquad d=1,\ldots,D.
    \label{eq:component_minmax}
\end{equation}

For works in the 1875--1940 fitting sample, each retained component has minimum zero and maximum one. 
When the same transformation is applied to works outside this period, the resulting values are clipped to the unit interval:
\begin{equation}
    u_{wd}
    =
    \min\left\{
        1,
        \max\left\{0,\tilde{z}_{wd}\right\}
    \right\}.
    \label{eq:component_clip}
\end{equation}
The value $u_{wd}$ is the final normalized work-level score used in the country-level aggregation. 
Normalization is performed separately for each PCA dimension and is not computed jointly across components or cultural groups.

\subsubsection{Step 4: Causal Kalman filtering}

For cultural group $i$, PCA dimension $d$, and calendar year $t$, let
$\mathcal{W}_i(t)$ denote the set of works assigned to group $i$ with
annotated composition year $t$. Country assignments account for a
composer's migration history when such information is available. If
$N_i(t)=|\mathcal{W}_i(t)|>0$, the annual observation is

\begin{equation}
    y_{id}(t)
    =
    \frac{1}{N_i(t)}
    \sum_{w\in\mathcal{W}_i(t)}
    u_{wd},
    \label{eq:annual_observation}
\end{equation}

where $u_{wd}$ is the normalized score of work $w$ on PCA dimension $d$.
If $N_i(t)=0$, the observation is treated as missing.

We model $y_{id}(t)$ as a noisy observation of a latent country-level
harmonic state $s_{id}(t)$. For each group--dimension pair, the state
follows a random-walk model:

\begin{align}
    s_{id}(t)
    &=
    s_{id}(t-1)+\eta_{id}(t),
    &
    \eta_{id}(t)
    &\sim
    \mathcal{N}(0,Q),
    \label{eq:kalman_state}
    \\
    y_{id}(t)
    &=
    s_{id}(t)+\epsilon_{id}(t),
    &
    \epsilon_{id}(t)
    &\sim
    \mathcal{N}\!\left(0,R_i(t)\right).
    \label{eq:kalman_measurement}
\end{align}

The process variance is fixed at $Q=0.0005$ for all cultural groups and
PCA dimensions. Because each annual observation is an average over
$N_i(t)$ works, we use the inverse-sample-size approximation

\begin{equation}
    R_i(t)
    =
    \frac{R_0}{N_i(t)},
    \qquad
    R_0=0.01.
    \label{eq:kalman_measurement_variance}
\end{equation}

Thus, annual observations based on more works are treated as more
reliable and receive greater weight in the filtering update. This
specification is a parsimonious approximation to the variance of a
sample mean; the empirical within-year variance is recorded for
diagnostic purposes but is not used in the measurement model.

For each group--dimension pair, the filter is initialized at the earliest
available observation year $t_{0,id}$:
$\hat{s}_{id}(t_{0,id}\mid t_{0,id})
    =
    y_{id}(t_{0,id}),
     \
    P_{id}(t_{0,id}\mid t_{0,id})
    =
    R_i(t_{0,id})$. The recursion is then run strictly forward in time. When an observation
is available, the state is updated according to its uncertainty
$R_i(t)$; when it is missing, only the prediction step is applied.
Consequently, $\hat{s}_{id}(t\mid t)$ depends only on observations available at or before year $t$.

For the subsequent dynamics analysis, the filtered annual trajectories
are sampled at five-year intervals. Let $\mathcal{T}=\{1875,1880,\ldots,1940\}$ denote the analysis grid. For each $t\in\mathcal{T}$, define $x_{id}(t)=\hat{s}_{id}(t\mid t)$ and collect the filtered states into
$\mathbf{X}(t)=\left[x_{id}(t)
    \right]_
    {i=1,\ldots,K;\,d=1,\ldots,D}
    \in
    \mathbb{R}^{K\times D}$. The matrix $\mathbf{X}(t)$ is the multidimensional harmonic-state
representation used as input to the DG and FJ models. PCA projection and component-wise scaling are estimated over
the full 1875--1940 analysis window to define a shared retrospective coordinate system, whereas the dynamics models are fitted on 1875--1925 and evaluated recursively on 1930--1940.

\begin{figure*}[t]
\centering
\includegraphics[scale=0.15]{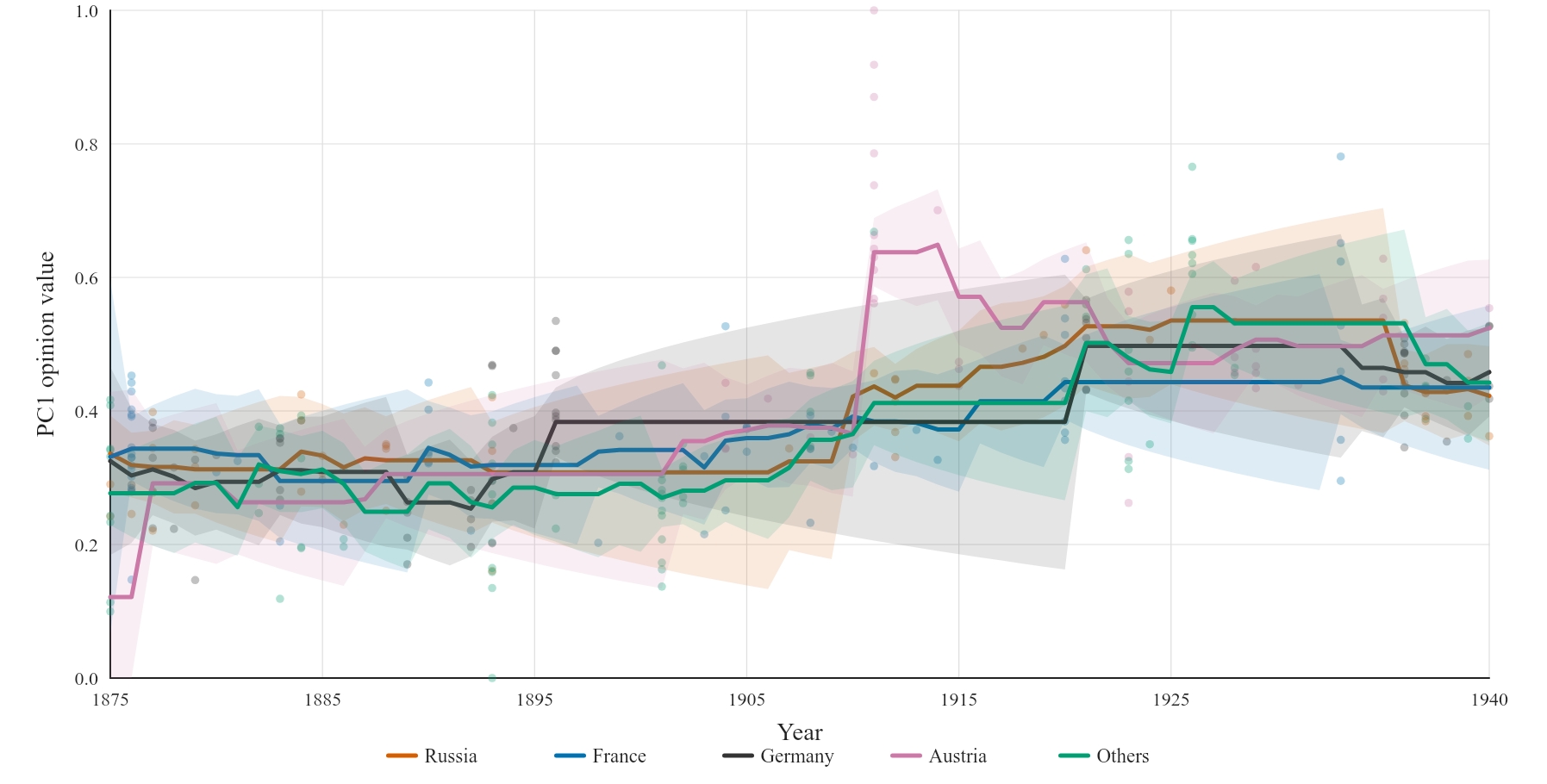}
\caption{Country-level opinion trajectories on PC1} \label{final_opinion}
\end{figure*}

For visualization, Fig.~\ref{final_opinion} shows the filtered country-level trajectories on PC1. Trajectories for all retained PCA dimensions are reported in Appendix~\ref{app:pca-trajectories}, Fig.~\ref{fig:all-pca-trajectories}.

\subsection{Opinion Dynamics Modeling}
\label{sec:opinion_dynamics}
\subsubsection{Multi-dimensional DG model}
\label{subsec:DG}

Let $\mathbf{X}(t)\in\mathbb{R}^{K\times D}$ collect the five countries'
Kalman-filtered harmonic opinion vectors at time $t$, with $D\leq 23$. Rows
correspond to country groups and columns correspond to retained PCA coordinates.
Thus, each country is represented by a $D$-dimensional harmonic state vector.

A general bilinear multidimensional transition could be written as
\begin{equation}
\mathbf{X}(t+1)
=\mathbf{C}\mathbf{W}\mathbf{X}(t),
\label{eq:general_multidim_transition}
\end{equation}
where $\mathbf{W}\in\mathbb{R}^{K\times K}$ models cross-country dependence and
$\mathbf{C}\in\mathbb{R}^{D\times D}$ models cross-component dependence among
PCA coordinates. In the main analysis, we use a restricted multidimensional
DG specification by setting $\mathbf{C}=\mathbf{I}_{D}$:
\begin{equation}
\mathbf{X}(t+1)=\mathbf{W}\mathbf{X}(t),
\qquad
\mathbf{W}\mathbf{1}=\mathbf{1},\quad 0\leq W_{ij}\leq1.
\label{eq:DG}
\end{equation}

Equivalently, for country $i$,
\begin{equation}
\mathbf{X}_i(t+1)
=
\sum_{j} W_{ij}\mathbf{X}_j(t),
\label{eq:DG_rowwise_vector}
\end{equation}
where $\mathbf{X}_i(t)\in\mathbb{R}^{D}$ is the harmonic state vector of country
$i$. This formulation is multidimensional in the sense that the opinion state
of each country is vector-valued, but the same country-level influence matrix
$\mathbf{W}$ is shared across all retained PCA coordinates and no lagged
cross-component transition is estimated.

Under this convention, rows of $\mathbf{W}$ correspond to target countries and
columns correspond to source countries. Therefore, $W_{ij}$ measures the
contribution of source country $j$ to the next-period harmonic state of target
country $i$. The restriction $\mathbf{C}=\mathbf{I}_{D}$ is a parsimonious modeling choice.
It does not imply that orthogonality alone rules out every possible lagged
cross-component effect; rather, it allows the same country-level influence
matrix to be estimated across all retained PCA dimensions. For later use, define the DG residual as
\begin{equation}
    \mathbf{R}_{\mathrm{DG}}(t;\mathbf{W})
    =
    \mathbf{X}(t+1)-\mathbf{W}\mathbf{X}(t).
    \label{eq:DG_residual}
\end{equation}

\subsubsection{FJ model}
\label{subsec:fj}

The multidimensional FJ (FJ) specification augments social
influence with anchoring to each country's intrinsic harmonic opinion matrix
$\mathbf{Z}\in\mathbb{R}^{K\times D}$:
\begin{equation}
    \mathbf{X}(t+1)=
    \mathbf{S}\mathbf{C}\mathbf{W}\mathbf{X}(t)\
    +(\mathbf{I}_K-\mathbf{S})\mathbf{Z},
    \label{eq:fj}
\end{equation}
where $\mathbf{C}=\mathbf{I}_{D}$, $\mathbf{Z}=\mathbf{X}(1875)$,
$\mathbf{S}=\operatorname{diag}(s_1,\ldots,s_K)$, and
$0\leq s_i\leq1$. Under this parameterization, $s_i$ is susceptibility to
social influence: $s_i=1$ gives the DG limit, whereas $s_i=0$ corresponds
to complete attachment to the intrinsic state. Hence, cultural stubbornness is
measured by $1-s_i$, not by $s_i$.

For stable estimation, we reparameterize the model in terms of the effective
influence matrix
\begin{equation}
    \mathbf{A}=\mathbf{S}\mathbf{W}, \qquad
    A_{ij}\geq0, \qquad
    \sum_j A_{ij}=s_i\leq1.
    \label{eq:fj_effective_matrix}
\end{equation}
Thus, $\mathbf{A}$ is a substochastic matrix whose row sum gives the
susceptibility of the corresponding country. This reparameterization converts
the bilinear parameterization in $(\mathbf{S},\mathbf{W})$ into a constrained
least-squares problem over $\mathbf{A}$. Equivalently, for target country $i$,
the FJ transition can be written as
\begin{equation}
    \mathbf{X}_i(t+1)
    =
    \mathbf{Z}_i
    +
    \sum_j A_{ij}\left[\mathbf{X}_j(t)-\mathbf{Z}_i\right],
    \label{eq:fj_rowwise}
\end{equation}
which is linear in the row $\mathbf{A}_i$ under the constraints
$A_{ij}\geq0$ and $\sum_j A_{ij}\leq1$.

The FJ residual is 
\begin{equation}
    \mathbf{R}_{\mathrm{FJ}}(t;\mathbf{A})
    =
    \mathbf{X}(t+1)
    -
    \mathbf{A}\mathbf{X}(t)
    -
    \left[\mathbf{I}-\operatorname{diag}(\mathbf{A}\mathbf{1})\right]\mathbf{Z}.
    \label{eq:fj_residual}
\end{equation}
After estimating $\widehat{\mathbf{A}}$, we recover
$\widehat{s}_i=\sum_j\widehat{A}_{ij}$ and, for rows with
$\widehat{s}_i>0$,$\widehat{W}_{ij}=\frac{\widehat{A}_{ij}}{\widehat{s}_i}.$
Thus, $\mathbf{W}$ contains the row-stochastic FJ influence weights, whereas
$\mathbf{A}=\mathbf{S}\mathbf{W}$ is the effective social influence operator
used in the transition equation.

\subsubsection{Equal-weight and PCA-variance-weighted objective functions}
\label{subsec:weighted_estimation}

Both the DG and FJ models are estimated on the same
five-country state matrices and the same 1875--1925 training period. The years
1930, 1935, and 1940 are held out and predicted recursively as one-, two-, and
three-step-ahead forecasts, respectively.

Let $\mathcal{T}_{\mathrm{train}}$ denote the set of training transitions, and
let $m\in\{\mathrm{DG},\mathrm{FJ}\}$ index the dynamical model. Each model is
specified by a parameter matrix $\boldsymbol{\theta}_m$, a residual function
$\mathbf{R}_m(t;\boldsymbol{\theta}_m)$, and a feasible parameter set
$\Theta_m$.

For the DG model, $\boldsymbol{\theta}_{\mathrm{DG}}=\mathbf{W}$
with feasible set 
    $\Theta_{\mathrm{DG}}
    =
    \left\{
    \mathbf{W}:\;
    \mathbf{W}\geq0,\;
    \mathbf{W}\mathbf{1}=\mathbf{1}
    \right\}$, And for the FJ model, we estimate the effective influence matrix
$\boldsymbol{\theta}_{\mathrm{FJ}}=\mathbf{A}=\mathbf{S}\mathbf{W}$ with feasible set
$\Theta_{\mathrm{FJ}}
    =
    \left\{
    \mathbf{A}:\;
    \mathbf{A}\geq0,\;
    \mathbf{A}\mathbf{1}\leq\mathbf{1}
    \right\}.$

To define equal-weight and PCA-variance-weighted estimation in a unified form,
let $q\in\{\mathrm{EW},\mathrm{VW}\}$ denote the estimation scheme. For the
equal-weight scheme, all retained PCA coordinates receive the same weight: $\mathbf{G}_D^{(\mathrm{EW})}=\mathbf{I}_D$.

For the PCA-variance-weighted scheme, let $v_d$ denote the explained-variance
ratio of PC$d$, and define
\begin{equation}
    \omega_d=\frac{v_d}{\sum_{k=1}^{D}v_k},
    \qquad
    \sum_{d=1}^{D}\omega_d=1,
    \qquad
    \boldsymbol{\Omega}_D=
    \operatorname{diag}(\omega_1,\ldots,\omega_D),
    \qquad
    \mathbf{G}_D^{(\mathrm{VW})}
    =
    \boldsymbol{\Omega}_D^{1/2}.
    \label{eq:pca_weights}
\end{equation}

The objective function for model $m$ under estimation scheme $q$ is
\begin{equation}
    \mathcal{L}_{m}^{(q)}(\boldsymbol{\theta}_m)
    =
    \sum_{t\in\mathcal{T}_{\mathrm{train}}}
    \left\|
    \mathbf{R}_m(t;\boldsymbol{\theta}_m)
    \mathbf{G}_D^{(q)}
    \right\|_F^2 .
    \label{eq:generic_objective}
\end{equation}
The corresponding parameter estimate is obtained by solving the constrained
optimization problem
\begin{equation}
    \widehat{\boldsymbol{\theta}}_{m}^{(q)}
    =
    \arg\min_{\boldsymbol{\theta}_m}
    \mathcal{L}_{m}^{(q)}(\boldsymbol{\theta}_m)
    \quad
    \text{s.t.}\quad
    \boldsymbol{\theta}_m\in\Theta_m.
    \label{eq:generic_optimization}
\end{equation}

Equivalently, if $\mathbf{r}_{m,d}(t;\boldsymbol{\theta}_m)$ denotes the
$d$-th column of the residual matrix
$\mathbf{R}_m(t;\boldsymbol{\theta}_m)$, then
\begin{equation}
    \mathcal{L}_{m}^{(q)}(\boldsymbol{\theta}_m)
    =
    \sum_{t\in\mathcal{T}_{\mathrm{train}}}
    \sum_{d=1}^{D}
    \alpha_d^{(q)}
    \left\|
    \mathbf{r}_{m,d}(t;\boldsymbol{\theta}_m)
    \right\|_2^2,
    \label{eq:generic_columnwise_objective}
\end{equation}
where
\begin{equation}
    \alpha_d^{(\mathrm{EW})}=1,
    \qquad
    \alpha_d^{(\mathrm{VW})}=\omega_d.
    \label{eq:objective_component_weights}
\end{equation}
Thus, the equal-weight objective assigns the same importance to every retained
PCA coordinate, whereas the variance-weighted objective gives greater weight to
coordinates explaining more embedding variance.

The same normalized PCA weights are used for both dynamical specifications.
Thus, differences between the equal-weight and variance-weighted estimators are
attributable solely to the relative importance assigned to the response
dimensions. PCA weighting modifies only the objective function, not the
transition equation. The DG forecasts are computed as $\widehat{\mathbf{X}}(t+1)
    =
    \widehat{\mathbf{W}}\mathbf{X}(t)$,
whereas the FJ forecasts are computed as
\begin{equation}
    \widehat{\mathbf{X}}(t+1)
    =
    \widehat{\mathbf{A}}\mathbf{X}(t)
    +
    \left[
    \mathbf{I}
    -
    \operatorname{diag}(\widehat{\mathbf{A}}\mathbf{1})
    \right]\mathbf{Z}.
    \label{eq:fj_forecast}
\end{equation}
In all cases, $\mathbf{C}=\mathbf{I}_D$. Setting
$\mathbf{C}=\boldsymbol{\Omega}_D$ would instead impose component-specific
attenuation at every recursive step and would define a different dynamical
model.

\subsubsection{Evaluation design and model comparison}
\label{subsec:evaluation}

All models are evaluated using the same chronological split. Model parameters
are estimated from the ten five-year transitions between 1875 and 1925. The
states in 1930, 1935, and 1940 are held out. Starting from the observed 1925
state, the three test states are forecast recursively: the 1930 prediction is
used as the input for 1935, and the 1935 prediction is used as the input for
1940. No observed test state is fed back into the model. The Kalman opinion
signals are computed by forward filtering, so the filtered state at each year
uses only observations available at that year or earlier. PCA projection and
component-wise normalization are defined over the full 1875--1940 analysis
window, providing a shared retrospective coordinate system for the evaluation.

Let $e_{tid}=\widetilde{X}_{id}(t)-X_{id}(t)$
denote a recursive-trajectory error, where $\widetilde{\mathbf{X}}(t)$ is the
model-generated state and $\mathbf{X}(t)$ is the filtered reference state. For
the training-trajectory metrics, the model is initialized once at the observed
1875 state and then recursively propagated through 1925 without being reset to
the observations. The initial 1875 error is therefore zero by construction.

To separate the parameter-estimation objective from the evaluation metric, each
fitted model is scored under two evaluation schemes. Let
$p\in\{\mathrm{EW},\mathrm{VW}\}$ index the evaluation scheme, and define
component weights
\begin{equation}
    \beta_d^{(\mathrm{EW})}=\frac{1}{D},
    \qquad
    \beta_d^{(\mathrm{VW})}=\omega_d,
    \qquad
    \sum_{d=1}^{D}\beta_d^{(p)}=1.
    \label{eq:evaluation_component_weights}
\end{equation}
where $v_d$ is the explained-variance ratio of PC$d$. The equal-weight scheme
therefore assigns the same evaluation weight to every retained coordinate,
whereas the variance-weighted scheme gives greater evaluation weight to
coordinates explaining more embedding variance.

Let $\mathcal{T}_{\mathrm{rec}}$ denote the set of recursive training states
used for trajectory evaluation, with $T=|\mathcal{T}_{\mathrm{rec}}|=11$. For
evaluation scheme $p$, the recursive MAE is
\begin{equation}
    \operatorname{MAE}^{(p)}
    =
    \sum_{d=1}^{D}\beta_d^{(p)}
    \frac{1}{TK}
    \sum_{t\in\mathcal{T}_{\mathrm{rec}}}
    \sum_{i=1}^{K}
    |e_{tid}|.
    \label{eq:evaluation_generic_mae}
\end{equation}
The mean trajectory RMSE is
\begin{equation}
    \operatorname{RMSE}_{\mathrm{mean}}^{(p)}
    =
    \sum_{d=1}^{D}\beta_d^{(p)}
    \frac{1}{K}
    \sum_{i=1}^{K}
    \left(
    \frac{1}{T}
    \sum_{t\in\mathcal{T}_{\mathrm{rec}}}
    e_{tid}^{2}
    \right)^{1/2}.
    \label{eq:evaluation_generic_rmse_mean}
\end{equation}
Thus, $\operatorname{RMSE}_{\mathrm{mean}}$ is the weighted mean of
country--component trajectory RMSE values; it is not a single pooled RMSE over
all entries.

For a held-out year $y\in\{1930,1935,1940\}$, the horizon-specific RMSE under
evaluation scheme $p$ is
\begin{equation}
    \operatorname{RMSE}_{y}^{(p)}
    =
    \left(
    \sum_{d=1}^{D}\beta_d^{(p)}
    \frac{1}{K}
    \sum_{i=1}^{K}
    e_{yid}^{2}
    \right)^{1/2}.
    \label{eq:evaluation_generic_horizon_rmse}
\end{equation}
For $p=\mathrm{EW}$, this reduces to the ordinary pooled cross-sectional RMSE
over the $K\times D$ entries of the held-out state. For $p=\mathrm{VW}$, the
same cross-sectional RMSE is weighted by PCA explained variance across
coordinates.

The training-residual column measures one-step in-sample fit rather than
recursive forecast performance. Let
\begin{equation}
    r_{tid}
    =
    f_{\widehat{\boldsymbol{\theta}}}\bigl(\mathbf{X}(t)\bigr)_{id}
    -
    X_{id}(t+5),
    \qquad
    t\in\mathcal{T}_{\mathrm{tr}},
    \label{eq:evaluation_onestep_residual}
\end{equation}
where the input at every training transition is the observed state and
$\mathcal{T}_{\mathrm{tr}}=\{1875,1880,\ldots,1920\}$ contains
$T_{\mathrm{tr}}=10$ training transitions. The one-step training RMSE under
evaluation scheme $p$ is
\begin{equation}
    \operatorname{TrainRMSE}^{(p)}
    =
    \left(
    \sum_{d=1}^{D}\beta_d^{(p)}
    \frac{1}{T_{\mathrm{tr}}K}
    \sum_{t\in\mathcal{T}_{\mathrm{tr}}}
    \sum_{i=1}^{K}
    r_{tid}^{2}
    \right)^{1/2}.
    \label{eq:evaluation_generic_train_rmse}
\end{equation}
This diagnostic is reported on the same RMSE scale as the recursive and
held-out error metrics.

In the model labels, ``equal'' and ``PCA weighted'' identify the objective used
for parameter estimation. In the evaluation panels, ``EQ'' and ``VW'' identify
how the fitted model is scored. These two choices are intentionally separated:
a model fitted with an equal-weight objective can be evaluated with
variance-weighted metrics, and a model fitted with a PCA-variance-weighted
objective can be evaluated with ordinary equal-coordinate metrics. PCA weighting
affects either the estimation objective or the evaluation metric, but it does
not change the transition equation or the restriction $\mathbf{C}=\mathbf{I}_D$.

Table~\ref{tab:model_comparison} applies this common protocol to DG and FJ. Evaluating every fitted specification under both metric systems
avoids mechanically favoring a PCA-weighted estimator merely because its
training loss used the same weights. Appendix~\ref{app:DG-weight-stability} shows that the estimated DG influence weights become comparatively stable as the number of retained PCA dimensions increases.

\begin{table*}[t]
\centering
\small
\caption{Model comparison at 23 PCA dimensions under ordinary and
PCA-variance-weighted evaluation. The first column reports the model family,
and the second column reports the estimation weighting scheme. ``PCA-weight''
refers to variance-weighted estimation. Lower is better. Bold denotes the best
performance within each panel.}
\label{tab:model_comparison}
\setlength{\tabcolsep}{4.5pt}
\renewcommand{\arraystretch}{1.08}
\begin{tabular}{llcccccc}
\toprule
\multirow{2}{*}{\textbf{Model}} &
\multirow{2}{*}{\textbf{Weight}} &
\multirow{2}{*}{\textbf{MAE}} &
\multicolumn{5}{c}{\textbf{RMSE}} \\
\cmidrule(lr){4-8}
& & &
\textbf{Mean} &
\textbf{1930} &
\textbf{1935} &
\textbf{1940} &
\textbf{Train} \\
\midrule

\multicolumn{8}{l}{\textit{Panel A: Ordinary equal-coordinate evaluation}} \\
\multirow{2}{*}{DG}
& Equal      & 0.0392 & 0.0498 & 0.0309          & 0.0469          & 0.0520          & 0.0420 \\
& PCA-weight & 0.0391 & 0.0498 & \textbf{0.0295} & \textbf{0.0466} & 0.0524          & 0.0422 \\
\addlinespace[2pt]
\multirow{2}{*}{FJ}
& Equal      & 0.0390          & 0.0496          & 0.0320 & 0.0479 & \textbf{0.0519} & \textbf{0.0419} \\
& PCA-weight & \textbf{0.0385} & \textbf{0.0492} & 0.0305 & 0.0471 & 0.0520          & 0.0421 \\

\midrule
\multicolumn{8}{l}{\textit{Panel B: PCA-variance-weighted evaluation}} \\
\multirow{2}{*}{DG}
& Equal      & 0.0458 & 0.0587 & 0.0308          & \textbf{0.0444} & 0.0525          & 0.0455 \\
& PCA-weight & 0.0449 & 0.0579 & \textbf{0.0298} & 0.0445          & 0.0529          & 0.0453 \\
\addlinespace[2pt]
\multirow{2}{*}{FJ}
& Equal      & 0.0458          & 0.0585          & 0.0329 & 0.0465 & 0.0522          & 0.0454 \\
& PCA-weight & \textbf{0.0442} & \textbf{0.0570} & 0.0313 & 0.0450 & \textbf{0.0519} & \textbf{0.0452} \\

\bottomrule
\end{tabular}
\end{table*}

\paragraph{Results under ordinary evaluation.}
PCA-weighted estimation improves the recursive MAE and mean RMSE of all three
model families, although the changes are small. The clearest improvement is for
FJ: its MAE decreases from $0.03903$ to $0.03854$, and its mean RMSE decreases
from $0.04964$ to $0.04922$, the best values in Panel A. PCA-weighted DG
provides the most accurate early forecasts, with RMSE $0.02945$ in 1930 and
$0.04659$ in 1935. At the longest horizon, however, equal-weight FJ is best
with RMSE $0.05190$. Equal-weight FJ also has the smallest one-step training
residual, $0.0419$. Its weighted counterpart sacrifices a small amount of
one-step fit but improves the recursively accumulated training errors. This
difference illustrates why a one-step residual and a recursive-path metric need
not rank models identically.

\paragraph{Results under PCA-variance-weighted evaluation.}
The advantage of weighted estimation becomes more visible when errors on the
leading principal components receive greater importance. PCA-weighted FJ has
the lowest MAE ($0.04424$), mean RMSE ($0.05703$), 1940 RMSE ($0.05192$), and
weighted residual ($0.10220$). PCA-weighted DG remains best for the 1930
forecast, with RMSE $0.02981$, whereas equal-weight DG is marginally best
in 1935, with RMSE $0.04435$ compared with $0.04445$ for weighted DG.
Consequently, variance weighting improves the overall representation of the
more informative PCA directions, but it does not guarantee lower error at every
individual forecast horizon.

\begin{figure*}[htbp]
    \centering
    \begin{minipage}{0.47\textwidth}
        \centering
        \includegraphics[width=\linewidth]{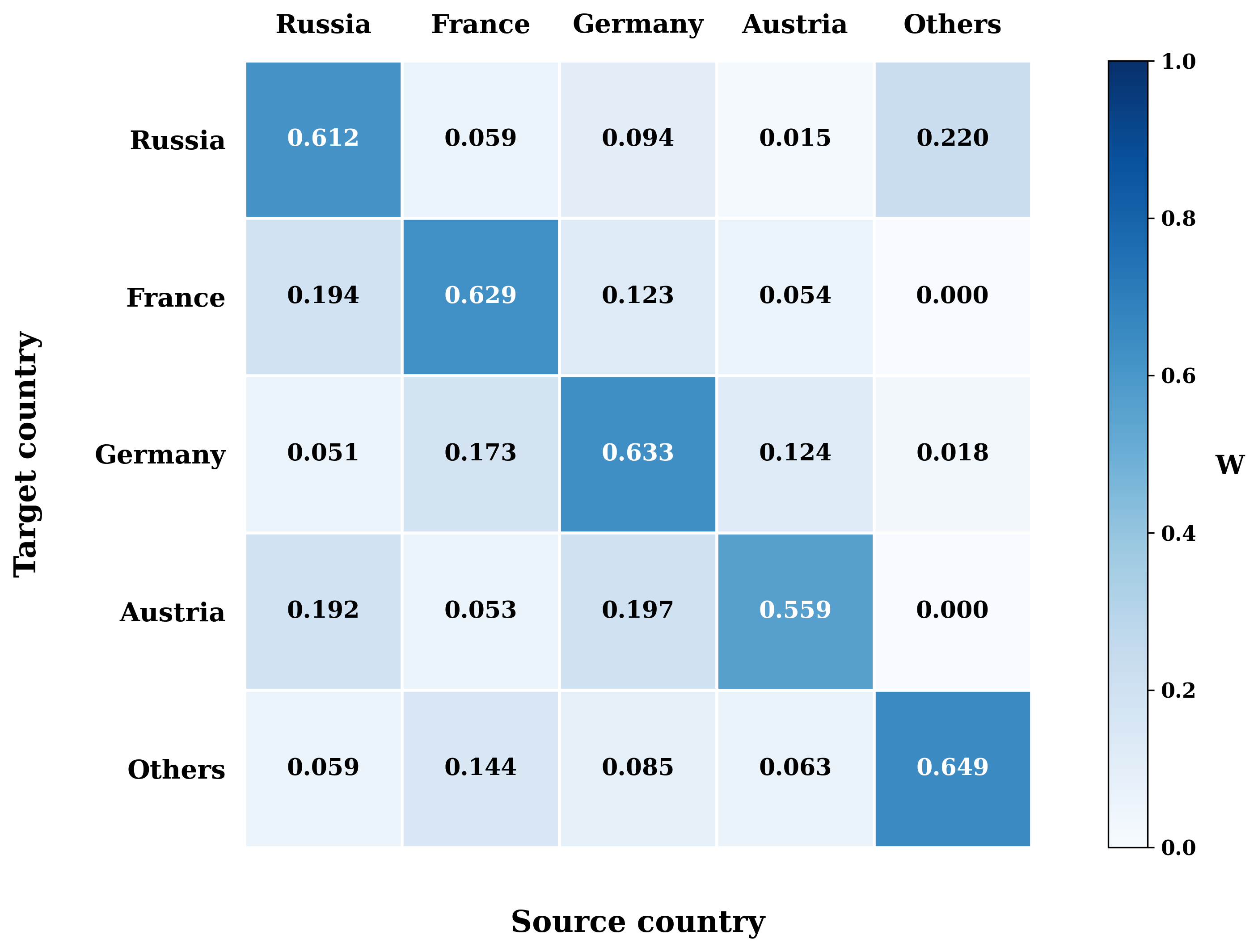}
        \\[2pt]\textbf{(a) DG}
    \end{minipage}
    \hfill
    \begin{minipage}{0.47\textwidth}
        \centering
        \includegraphics[width=\linewidth]{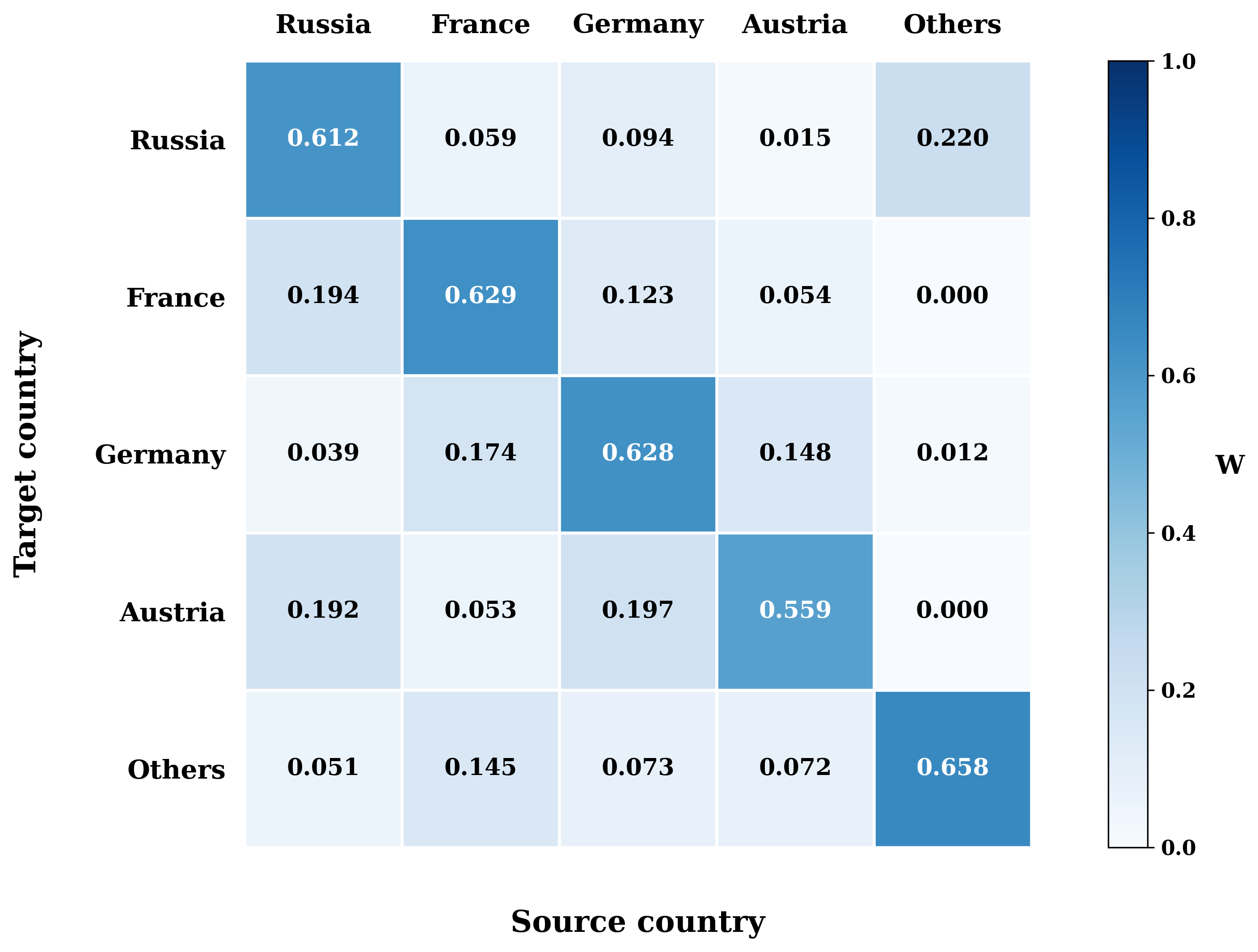}
        \\[2pt]\textbf{(b) FJ}
    \end{minipage}
    \caption{Estimated DG influence matrix and FJ
    row-stochastic influence weights from the 23-dimensional models. Rows
    denote target countries and columns denote source countries. The FJ panel
    displays the row-stochastic matrix $\mathbf{W}$ before susceptibility
    scaling; its effective social influence operator is
    $\mathbf{S}\mathbf{W}$.}
    \label{fig:heatmap_comparison}
\end{figure*}

\subsubsection{Parameter interpretation}
\label{subsec:parameter_interpretation}

Fig.~\ref{fig:heatmap_comparison} shows the estimated country-level dependency
matrices for the 23-dimensional models with PCA variance weighted. Rows denote target countries and columns
denote source countries, so an entry in row $i$ and column $j$ represents the
contribution of source country $j$ to the next-period harmonic state of target
country $i$. The DG panel displays the estimated row-stochastic matrix
$\widehat{\mathbf{W}}$. The FJ panel displays the row-stochastic
matrix $\widehat{\mathbf{W}}$ recovered from the effective matrix
$\widehat{\mathbf{A}}=\widehat{\mathbf{S}}\widehat{\mathbf{W}}$; the effective
social influence operator in the FJ transition is therefore
$\widehat{\mathbf{S}}\widehat{\mathbf{W}}$.

The diagonal entries of $\widehat{\mathbf{W}}$ summarize self-persistence in
each cultural trajectory. In the DG model, Russia, France, Germany,
Austria, and Others have self-dependency weights of approximately $0.612$,
$0.629$, $0.633$, $0.559$, and $0.649$, respectively. This indicates that each
group's next-period harmonic state is explained primarily by its own previous
state, rather than by immediate replacement through external sources. This
pattern is consistent with the persistence of national and institutional
traditions during the transition from late Romanticism to early modernism:
German and Austrian harmonic practices remained connected to the Austro-German
lineage of late Romantic chromaticism and Viennese modernism, French harmonic
development retained a distinct trajectory from Impressionism toward
neoclassicism, and Russian music combined national-school inheritances with
modernist
\cite{dahlhaus1989nineteenth,cook2004cambridge,griffiths2011modern}.

The off-diagonal entries reveal the main cross-group dependencies estimated by
the model. The DG matrix recovers a close German--Austrian relationship:
Austria assigns a substantial weight to Germany, while Germany also assigns a
non-negligible weight to Austria. This is historically plausible given the
shared Austro-German harmonic environment linking late Romantic idioms, Mahler's
expanded tonal language, and Viennese modernism \cite{shaw2010cambridge}. The
French and German trajectories also show asymmetric cross-dependence, reflecting
a broader European context in which French harmonic color, modal sonority, and
tonal ambiguity developed in dialogue with the dominant German Romantic and
modernist traditions.

The Russian-related dependencies are also historically meaningful. The Russian
row shows strong self-persistence, a relatively large dependency on the
composite Others group, and smaller dependencies on Germany and France. This
pattern is consistent with the mixed position of Russian music, which developed
from national-school traditions while also interacting with broader European
modernist currents through figures such as Scriabin, Stravinsky, Prokofiev, and
émigré composer networks \cite{frolova2007russian,taruskin1996stravinsky}.
Because the Others group combines Czech, Hungarian, Italian, Nordic, American,
and other repertoires, the Russia--Others dependency should be interpreted as a
broad aggregate signal rather than as a single coherent channel of influence.

$\widehat{W}_{\mathrm{France},\mathrm{Russia}}$
captures the contribution of the Russian trajectory to the next-period French
harmonic state. This value is approximately $0.194$ in both the DG and FJ
matrices, making Russia the largest external source in the French row after
France's own self-dependency
($\widehat{W}_{\mathrm{France},\mathrm{France}}\approx 0.629$). This is
consistent with the strong presence of Russian modernism in Parisian musical
life, especially through Diaghilev's Ballets Russes and the reception of
Stravinsky. The Paris premiere of \emph{Le Sacre du printemps} in 1913 provides
a salient example of how Russian-derived modernist materials entered the French
cultural environment \cite{taruskin1996stravinsky}. 

The Others group has the largest self-weight in the DG matrix. This does
not imply that the grouped countries form a unified musical tradition. Rather,
it reflects the fact that the composite category contains internally
heterogeneous materials whose averaged trajectory is partly stabilized by its own
previous state. Its weights are therefore useful for improving temporal coverage
and detecting broad non-core stylistic movement, but they should not be given
the same musicological interpretation as the rows corresponding to Russia,
France, Germany, or Austria.

The FJ matrix $\widehat{\mathbf{W}}$ should be interpreted differently from the
DG matrix. In DG, $\widehat{\mathbf{W}}$ is the full transition
operator. In FJ, $\widehat{\mathbf{W}}$ describes the composition of the social
component conditional on susceptibility, while the total strength of that
component is controlled by $\widehat{\mathbf{S}}$. Thus, the effective
contribution of source $j$ to target $i$ is
$\widehat{s}*i\widehat{W}*{ij}$, not simply $\widehat{W}_{ij}$.

For the FJ model, the equal-weight susceptibility estimates are
$\widehat{\mathbf{S}}=
(1.0000,1.0000,0.9536,1.0000,0.8399)$,
in the order Russia, France, Germany, Austria, and Others. The corresponding
stubbornness values, defined as $1-s_i$, are
$(0,0,0.0464,0,0.1601)$. Thus, under equal weighting, the Others group
shows the strongest estimated attachment to its intrinsic harmonic state, which
is consistent with its role as a heterogeneous aggregate.

Under PCA-variance weighting, the susceptibility estimates are $\widehat{\mathbf{S}}^{(\mathrm{VW})}=
(1.0000,1.0000,0.8900,1.0000,0.9263$). The corresponding stubbornness values are
$(0,0,0.1100,0,0.0737)$. In this specification, the strongest estimated
anchoring shifts from Others to Germany, suggesting that the leading PCA
dimensions emphasize harmonic variance associated with the German trajectory.

Overall, the estimated matrices suggest three broad historical patterns: strong
within-tradition persistence, a close German--Austrian dependency, and selective
cross-dependence among the Russian, French, and broader European trajectories.
These patterns are consistent with the historically dense transition from late
Romanticism to early modernism between 1875 and 1940.

\section{Discussion}
\label{sec:discussion}

\subsection{Music-Historical Reading of the Estimated Dynamics}

The estimated DG and FJ models should be read as compact summaries of harmonic co-evolution, not as direct evidence of cultural causation. Several patterns are compatible with established accounts of the late-Romantic and early-modernist transition. Strong self-dependence reflects the persistence of national and institutional traditions across five-year intervals. The close German--Austrian relationship is consistent with their shared Austro-German harmonic, pedagogical, and institutional environment. The Russian and French trajectories also support a historically plausible reading of asymmetric exchange: Russian harmonic practice developed in dialogue with Central European models, while Russian musical currents were also present in French modernist culture, particularly through reception networks and Paris-centered artistic activity. These patterns should be understood as model-based evidence of stylistic co-movement, not as proof of one-way influence among individual composers.

This reading is also supported, in a limited diagnostic sense, by the supplementary MIDI stress test reported in Appendix~\ref{app:midi_bilstm}. That experiment uses symbolic MIDI data and BiLSTM era-classifier probabilities rather than the ChordBERT/PCA harmonic representation used in the main analysis. The resulting influence matrix reproduces several coarse motifs of the main analysis, including strong within-agent persistence and broadly similar cross-group dependency patterns. This qualitative overlap suggests that some high-level features of the estimated dynamics are not entirely tied to a single representation pipeline.

\subsection{Aggregation and the ``Others'' Category}

The composite ``Others'' agent is useful for maintaining temporal coverage, but it is not a coherent cultural actor. In this corpus, it groups less densely represented traditions associated with Norway, Finland, Czechia, Italy, Hungary, the United States, and other contexts. These traditions are not marginal in historical or aesthetic terms; the problem is that a single state vector compresses several distinct national, regional, and stylistic trajectories.

This aggregation can affect both the estimated influence weights and the FJ anchoring terms. Apparent stubbornness, instability, or anomalous cross-dependencies in the ``Others'' row may partly reflect internal heterogeneity rather than a meaningful collective stylistic position. This limitation points to an important direction for future work: when sufficient data are available, the aggregate comparison group should be decomposed into finer national, regional, institutional, or stylistic agents.

\subsection{Parsimony and Interpretability}

The use of a shared influence matrix across PCA dimensions is a deliberate compromise between musical richness and model interpretability. A separate matrix for each harmonic dimension could capture feature-specific diffusion patterns, but the available historical time series is too short to support such a parameter-rich specification reliably. It would also make the resulting network difficult to interpret as a single cultural dependency structure.

The equal-coordinate and PCA-variance-weighted estimators therefore serve complementary roles. The weighted estimator emphasizes the leading harmonic dimensions and yields modest predictive gains, while the equal-coordinate estimator provides a transparent robustness check. Because both estimators preserve the same transition equation and the same single-matrix interpretation, their agreement supports the stability of the main qualitative conclusions without overstating the precision of the inferred influence network. The MIDI-based stress test extend this robustness discussion in two complementary directions: the former probes temporal variation in the dependency structure, while the latter tests whether similar coarse patterns can appear in a different symbolic representation space. Together, these diagnostics support the portability of the representation-to-dynamics framework while reinforcing the need for cautious historical interpretation.

\subsection{Limitations}

Several limitations should be emphasized. First, the main DG and FJ specifications use linear transitions and estimate a single dependency structure over the full analysis window. This choice supports interpretability, but it also compresses historically uneven processes into a long-run average. Events such as World War I, the Russian Revolution, institutional displacement, and composer migration may have altered the channels of musical exchange in ways that a fixed matrix cannot fully represent. historical change.

Second, the inferred states are harmonic-style trajectories rather than complete descriptions of musical style. They are derived from symbolic chord sequences and contextual chord embeddings, and therefore do not directly capture rhythm, timbre, orchestration, form, text setting, performance practice, or reception history. The supplementary MIDI experiment in Appendix~\ref{app:midi_bilstm} obtains a broadly similar coarse influence matrix from a different symbolic representation space. Since MIDI-based inputs can preserve rhythmic and temporal aspects that are absent from chord-label sequences, this result suggests that the proposed pipeline is not restricted to the ChordBERT/PCA harmonic representation and can be transferred to richer symbolic music inputs.

Third, the country-level trajectories depend on sparse and unevenly distributed historical observations. Causal Kalman filtering reduces noise and accounts for different annual sample sizes through measurement uncertainty, but it cannot recover information from genuinely missing periods. Trajectories based on fewer works remain more uncertain, especially for less densely represented traditions and for boundary regions of the analysis window. 

Fourth, the PCA projection and component-wise normalization are appropriate for retrospective reconstruction of the 1875--1940 period, but they do not constitute a fully leakage-free forecasting design. A stricter predictive setup would fit dimensionality reduction and normalization parameters only on the training period before applying them to later works. The present study should therefore be understood as historical modeling and interpretable retrospective analysis rather than real-time prediction.

Finally, the estimated matrices should not be interpreted as direct evidence that one country influenced another. They identify the dependencies that best explain the observed harmonic trajectories under the assumptions of the model. Historical mechanisms such as pedagogy, migration, publication, performance, institutional affiliation, and aesthetic opposition require additional archival and musicological evidence. The value of the framework is to provide a reproducible quantitative layer that can generate, support, or challenge historically grounded interpretations, not to replace qualitative music history.

\section{Conclusion}
\label{sec:conclusion}

This article has proposed a representation-to-dynamics framework for studying long-horizon stylistic change in symbolic musical heritage data. By linking chord-based representation learning, temporally causal trajectory reconstruction, and opinion-dynamics modeling, the framework transforms sparse work-level observations into interpretable group-level harmonic trajectories. In this formulation, national or cultural style is not treated as a fixed label, but as an evolving state whose movement can be compared across historical time.

The case study on Western art music from 1875--1940 shows that this framework can produce historically readable summaries of harmonic co-evolution. The estimated dynamics capture persistent self-dependence, close German--Austrian proximity, and plausible cross-currents among Russian, French, and Central European harmonic idioms. These patterns are broadly consistent with established accounts of late-Romantic and early-modernist musical exchange, while also showing how sparse observations, heterogeneous aggregation, and model specification choices affect the resulting influence-like matrices. The PCA-variance-weighted estimator offers modest predictive gains without changing the interpretation of a single cross-group dependency structure.

The supplementary diagnostics further clarify the scope of the approach. The MIDI-based stress test shows that the pipeline can be transferred to a different symbolic representation space that includes temporal and rhythmic information beyond chord labels. 

Future work should connect these dynamical summaries more closely to historical mechanisms, including institutions, pedagogy, migration, publication, performance, and reception. It should also extend the representation layer beyond harmony and refine the agent structure beyond coarse national groupings. More broadly, the value of the proposed framework lies in providing a reproducible quantitative layer for cultural heritage research: one that can make assumptions explicit, compare stylistic trajectories across time, and complement, rather than replace, archival and musicological interpretation.

\bibliographystyle{plain}
\bibliography{references}

\appendix
\label{appendix}

\section{Representation Evaluation Metrics}
\label{app:representation-metrics}

This appendix summarizes the evaluation metrics used for chord-level
masked-token prediction and work-level retrieval evaluation. Table~\ref{tab:evaluation-metrics}
provides the metric definitions and their corresponding optimization
directions.

Let $\mathcal{U}$ denote the set of masked chord-token instances and
$\mathcal{Q}$ denote the set of retrieval queries. For a masked instance
$u$, let $\rho_u^{\mathrm{tok}}$ denote the rank of the ground-truth
chord token. For a retrieval query $q$, let $y_{q,k}$ indicate whether
the item ranked at position $k$ is relevant, and let
$\rho_q^{\mathrm{ret}}$ denote the rank of the first relevant item.
This appendix provides formal definitions of the chord-level and
work-level metrics used to evaluate the representation models. Lower
values are better for masked-language-modeling loss, whereas higher
values are better for all ranking and retrieval metrics.
\begin{table}[t]
\centering
\caption{Evaluation metrics used for representation evaluation.}
\label{tab:evaluation-metrics}
\begin{tabular}{p{2.2cm} p{4.5cm} p{4.2cm}}
\toprule
Level & Metric & Definition \\
\midrule

Chord &
MLM Loss &
$
-\frac{1}{|\mathcal U|}
\sum_{u\in\mathcal U}
\log p_\theta(c_u|\tilde{\mathbf c}_u)
$
\\

&
Top-$K$ Accuracy &
$
\frac{1}{|\mathcal U|}
\sum_{u\in\mathcal U}
\mathbb{I}
[\rho_u^{\mathrm{tok}}\leq K]
$
\\

&
MRR &
$
\frac{1}{|\mathcal U|}
\sum_{u\in\mathcal U}
\frac{1}{\rho_u^{\mathrm{tok}}}
$
\\

\midrule

Work &
Recall@$K$ &
$
\frac{1}{Q}
\sum_{q\in\mathcal Q}
\mathbb{I}
[
\sum_{k=1}^{K}y_{q,k}>0
]
$
\\

&
MRR &
$
\frac{1}{Q}
\sum_{q\in\mathcal Q}
\frac{1}{\rho_q^{\mathrm{ret}}}
$
\\

&
mAP@$K$ &
$
\frac{1}{Q}
\sum_{q\in\mathcal Q}
AP@K_q
$
\\

\bottomrule
\end{tabular}
\end{table}

\subsection{Retrieval protocols}

In same-work movement retrieval, each movement serves as a query and
the remaining movements belonging to the same composition form its
relevant set $\mathcal{R}_q$. The query movement itself is excluded
from the candidate set. Because a multi-movement work may contain more
than one additional movement, this protocol supports both first-match
metrics, such as Recall@$K$ and MRR, and multi-relevant-item metrics,
such as mAP@10.

In split-half identity retrieval, the first half of each work is used as
a query and the corresponding second half is its only relevant item.
Thus, $R_q=1$ for every query. In this setting, Recall@$K$ indicates
whether the matching half appears within the first $K$ positions, while
MRR summarizes its exact rank.

\section{Supplementary Diagnostics for PCA Opinion States}
\label{app:pca-opinion-diagnostics}

This appendix provides supplementary diagnostic visualizations for the multidimensional harmonic opinion states used in the dynamics models. The main text reports representative trajectories and aggregate model performance, while the following figures make the underlying PCA-coordinate trajectories and model-level prediction behavior more transparent.

\subsection{Component-wise PCA Opinion Trajectories}
\label{app:pca-trajectories}

\begin{figure*}[t]
\centering
\includegraphics[width=\textwidth]{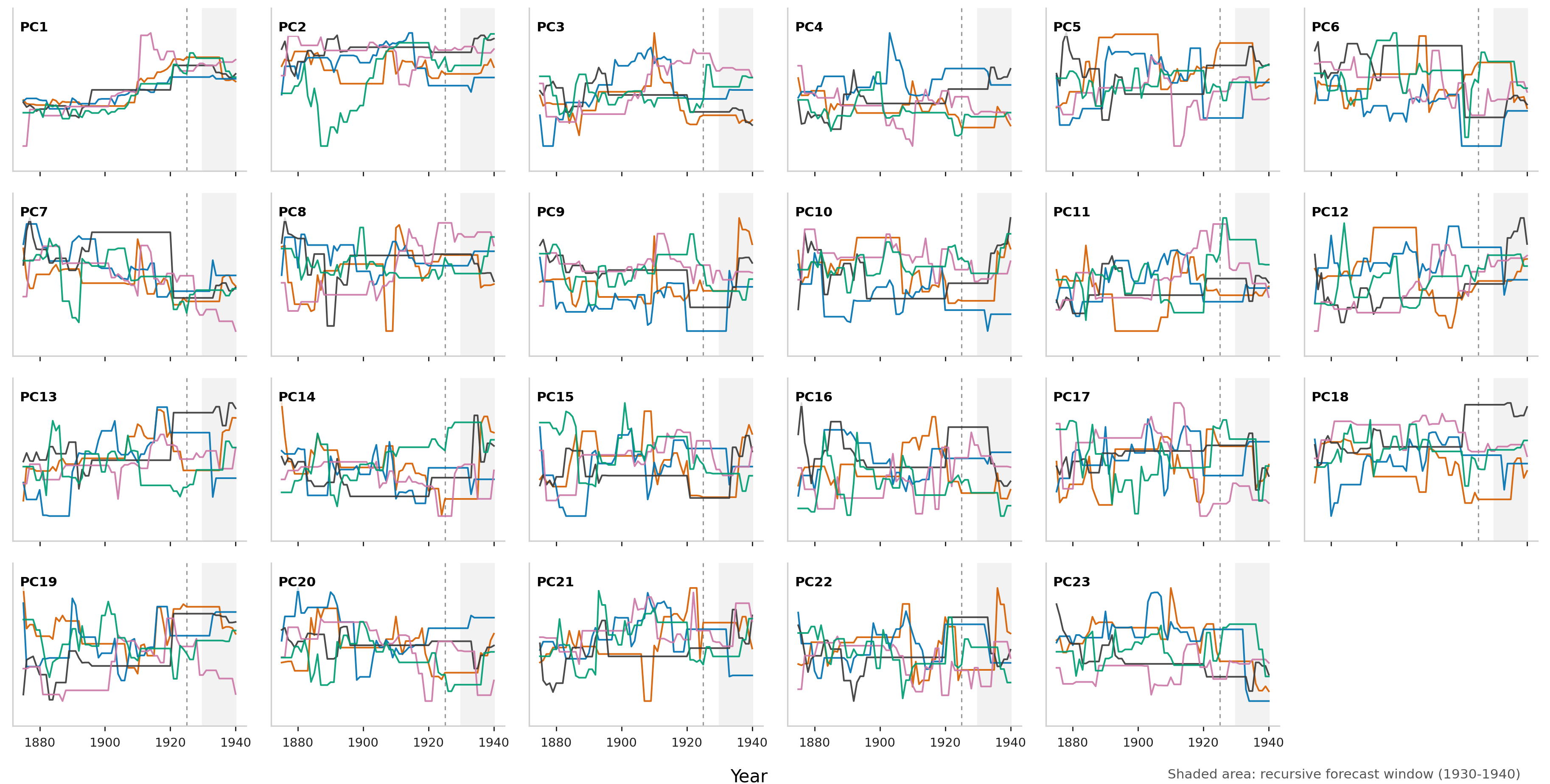}
\caption{
Component-wise country-level harmonic opinion trajectories for all retained PCA dimensions.
}
\label{fig:all-pca-trajectories}
\end{figure*}

To provide a complete view of the multidimensional state representation, Fig.~\ref{fig:all-pca-trajectories} reports the country-level harmonic opinion trajectories for all retained PCA dimensions. Each subplot corresponds to one PCA coordinate, and each curve shows the Kalman-filtered normalized opinion value of one cultural agent. The vertical dashed line marks the end of the training period in 1925, while the shaded region indicates the recursive forecast period from 1930 to 1940.

\subsection{Influence-Weight Stability Across PCA Dimensionality}
\label{app:DG-weight-stability}

\begin{figure*}[t]
    \centering
    \includegraphics[width=\textwidth]{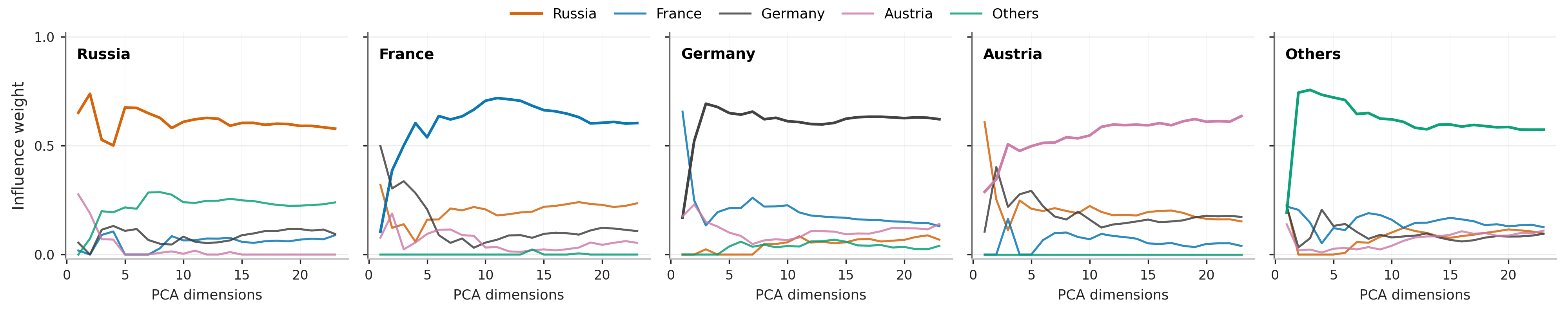}
    \caption{DG influence weights across retained PCA dimensionality under equal-coordinate estimation.}
    \label{fig:DG-weights-equal-pca}
\end{figure*}

\begin{figure*}[t]
    \centering
    \includegraphics[width=\textwidth]{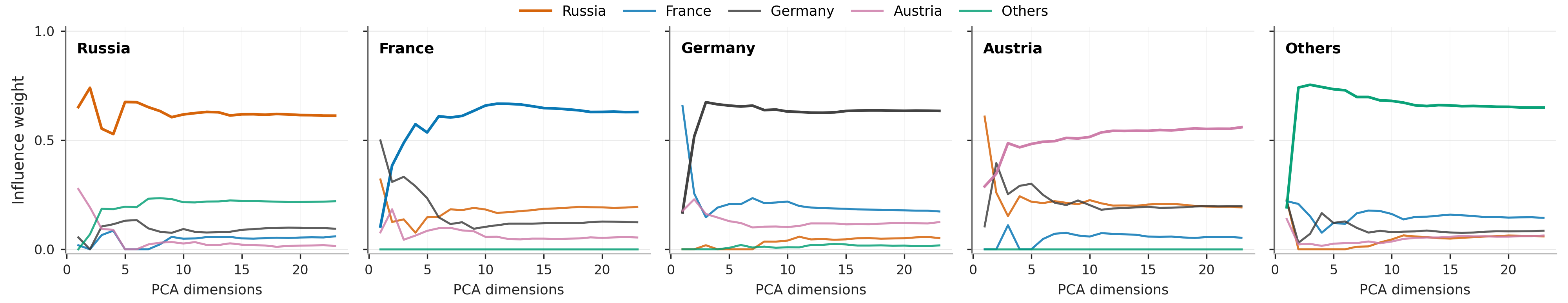}
    \caption{DG influence weights across retained PCA dimensionality under PCA-variance-weighted estimation.}
    \label{fig:DG-weights-weighted-pca}
\end{figure*}

To assess whether the estimated influence structure is sensitive to the number of retained PCA coordinates, Figs.~\ref{fig:DG-weights-equal-pca} and~\ref{fig:DG-weights-weighted-pca} report the DG influence weights as the number of retained leading PCA coordinates increases from 1 to 23. The equal-coordinate and PCA-variance-weighted estimators are shown separately. Each panel fixes a target country, and each curve gives the estimated contribution of one source country to that target country's next-period harmonic opinion state.

\subsection{PC1 Forecast Diagnostics Across Dynamics Models}
\label{app:pc1-forecast-diagnostics}

In addition to the component-wise opinion trajectories, we provide a direct visualization of recursive prediction behavior on the first principal component. Since PC1 accounts for the largest fraction of variance in the retained harmonic representation space, it offers a compact diagnostic view of how different dynamics models propagate the country-level opinion states over time.

\begin{figure*}[t]
\centering
\includegraphics[width=\textwidth]{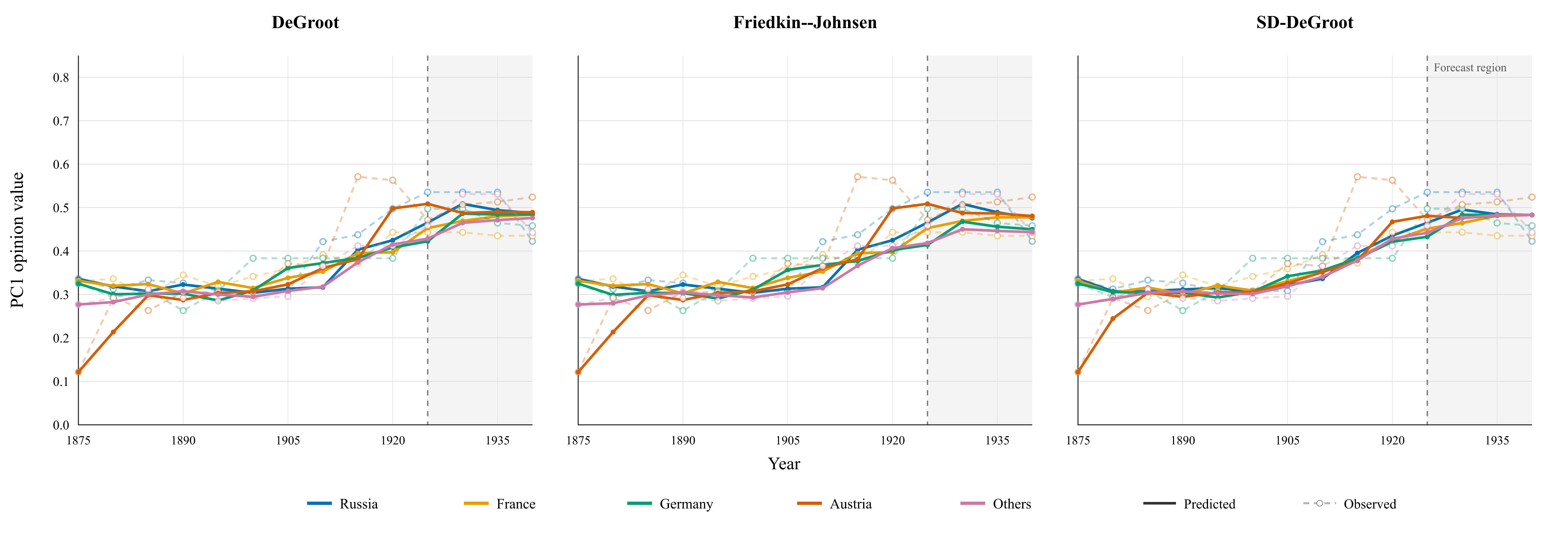}
\caption{
PC1 recursive prediction diagnostics for three dynamics models.
}
\label{fig:pc1-model-comparison}
\end{figure*}

Figure~\ref{fig:pc1-model-comparison} compares the PCA-variance-weighted DG model, the FJ model, and the state-dependent DG extension on PC1. Solid curves denote recursively predicted trajectories, while dashed curves denote the corresponding observed Kalman-filtered states. The vertical dashed line marks the end of the training period in 1925, and the shaded region denotes the recursive forecast period.

\section{MIDI BiLSTM Stress Test}
\label{app:midi_bilstm}

We conduct an additional symbolic-MIDI stress test to examine whether the
proposed state-construction and opinion-dynamics pipeline can be transferred to
a different work-level representation. This experiment uses the
\texttt{TiMauzi/imslp-midi-by-sa} MIDI corpus and the
\texttt{TiMauzi/EraClassifierBiLSTM-134M} model. In contrast to the main
ChordBERT/PCA harmonic state space, the MIDI experiment uses six supervised
era-classifier probabilities directly as opinion dimensions: Renaissance,
Baroque, Classical, Romantic, Modern, and Other.

Each MIDI file is converted to the model-card feature representation and passed
through the BiLSTM classifier. Window-level logits are averaged within each work
and then converted to probabilities by a softmax. Country-level states are
constructed with the same causal Kalman filtering procedure used in the main
analysis, using five-year bins, $Q=0.0005$, and $R_t=0.01/N_t$.

\begin{table}[t]
\centering
\caption{Summary of the MIDI BiLSTM stress-test dataset and opinion space.}
\label{tab:midi_bilstm_summary}
\begin{tabular}{lp{0.62\linewidth}}
\toprule
\textbf{Item} & \textbf{Value} \\
\midrule
Analysis window & 1875--1940 \\
Selected MIDI records & 778 records by 244 composers \\
Country resolution & 670 matched works; 108 unmatched works \\
Opinion dimensions & Renaissance, Baroque, Classical, Romantic, Modern, Other \\
Predicted argmax classes & Modern: 761; Renaissance: 15; Other: 2; Baroque/Classical/Romantic: 0 \\
Five-agent grouping & Russia: 26; France: 70; Germany: 106; Austria: 11; Others: 457 \\
Top-agent grouping & USA: 199; Germany: 106; Belgium: 75; France: 70; England: 59; Italy: 58; Russia: 26; Poland: 25; Others: 52 \\
\bottomrule
\end{tabular}
\end{table}

The main diagnostic finding is that the BiLSTM probability space collapses
toward the Modern class. As summarized in
Table~\ref{tab:midi_bilstm_summary}, 761 of the 778 MIDI records have Modern as
their predicted argmax class, while Baroque, Classical, and Romantic never appear
as the highest-probability class. This concentration may reflect the classifier,
the MIDI feature reconstruction, or the selected corpus itself. Therefore, this
experiment is treated as a stress test rather than as an alternative historical
opinion space.

\begin{figure}[t]
\centering
\includegraphics[width=\textwidth]{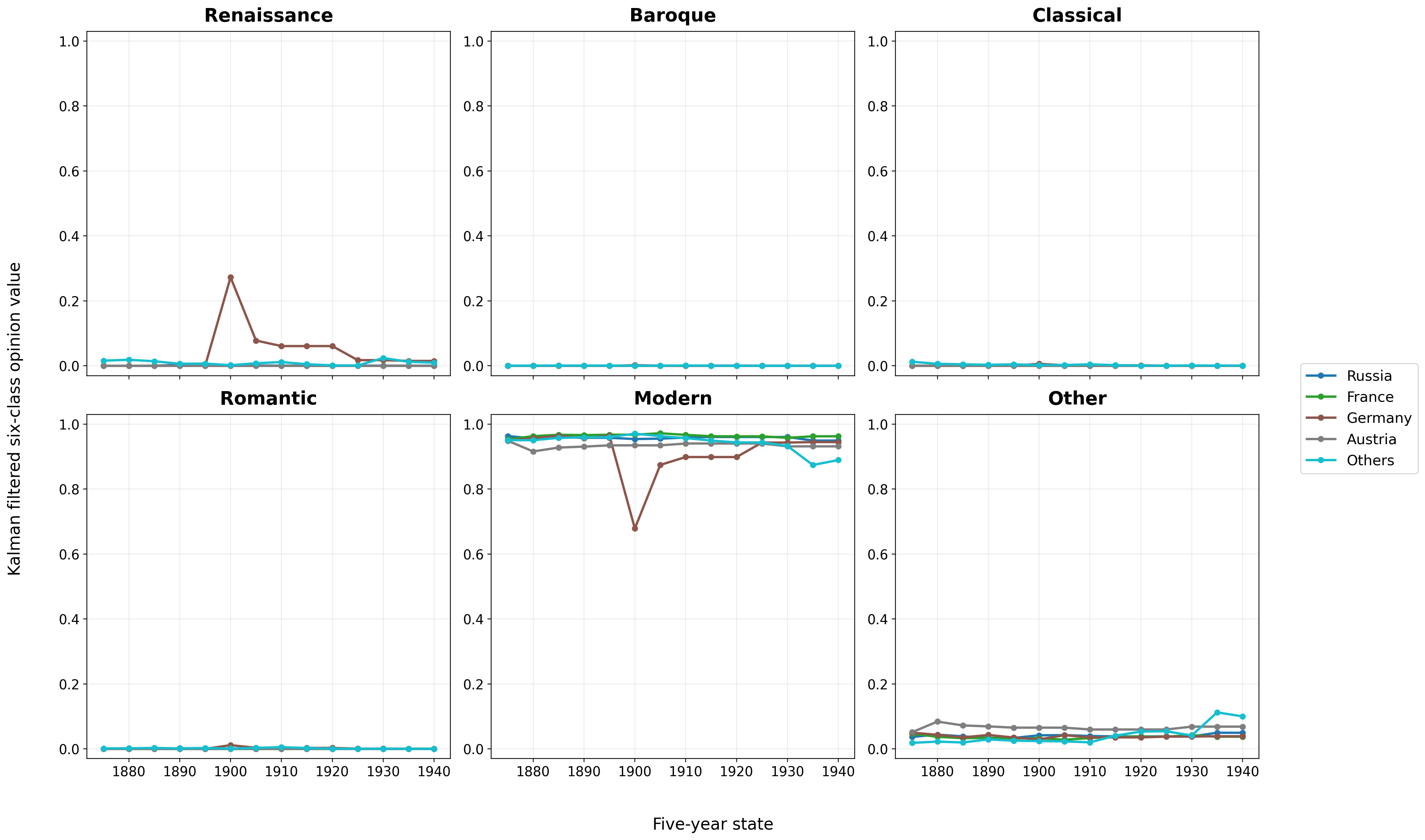}
\caption{Five-agent causal Kalman trajectories in the six-dimensional BiLSTM
era-probability opinion space. The Modern dimension dominates the trajectories,
while most other era dimensions remain close to zero.}
\label{fig:midi_five_agent_trajectories}
\end{figure}

Figure~\ref{fig:midi_five_agent_trajectories} visualizes the resulting
five-agent Kalman trajectories. The figure confirms that the opinion space is
dominated by the Modern dimension, with most other era-probability dimensions
remaining close to zero across countries and time. This motivates our decision
to interpret the MIDI experiment as a diagnostic check rather than as a direct
substitute for the ChordBERT/PCA representation used in the main analysis.

We evaluate both the original five-agent design and a MIDI-specific top-agent
design. The five-agent design preserves comparability with the main experiment,
whereas the top-agent design better reflects the country distribution of the
MIDI corpus. Models are fitted on 1875--1925 and recursively evaluated on 1930,
1935, and 1940. Table~\ref{tab:midi_bilstm_performance} reports performance
using the full six-dimensional probability vector.

\begin{table}[t]
\centering
\caption{Forecast performance for the MIDI BiLSTM stress test using the full
six-dimensional probability vector.}
\label{tab:midi_bilstm_performance}
\begin{tabular}{llrrrrr}
\toprule
\multirow{2}{*}{\textbf{Agent design}} &
\multirow{2}{*}{\textbf{Model}} &
\multirow{2}{*}{\textbf{MAE}} &
\multicolumn{4}{c}{\textbf{RMSE}} \\
\cmidrule(lr){4-7}
& & &
\textbf{mean} &
\textbf{1930} &
\textbf{1935} &
\textbf{1940} \\
\midrule
\multirow{2}{*}{Five-agent}
& DG & 0.005967 & 0.014660 & 0.006775 & 0.018722 & 0.015758 \\
& FJ      & 0.005981 & 0.014668 & 0.006777 & 0.018731 & 0.015769 \\
\midrule
\multirow{2}{*}{Top-agent}
& DG & 0.003760 & 0.010887 & 0.005148 & 0.014062 & 0.011461 \\
& FJ      & 0.003760 & 0.010806 & 0.005111 & 0.014037 & 0.011275 \\
\bottomrule
\end{tabular}
\end{table}

Although the absolute RMSE values are small, the fitted opinion-dynamics models
do not outperform a persistence baseline. For the five-agent design, the RMSE
skill values are $-0.124207$ for DG and $-0.124845$ for FJ. For the
top-agent design, the corresponding values are $-0.176693$ and $-0.167879$.
Thus, the high apparent trajectory similarity mainly reflects the near-constant
structure induced by the collapsed classifier-probability space.

\begin{figure}[t]
\centering
\begin{subfigure}{0.47\textwidth}
    \centering
    \includegraphics[width=\textwidth]{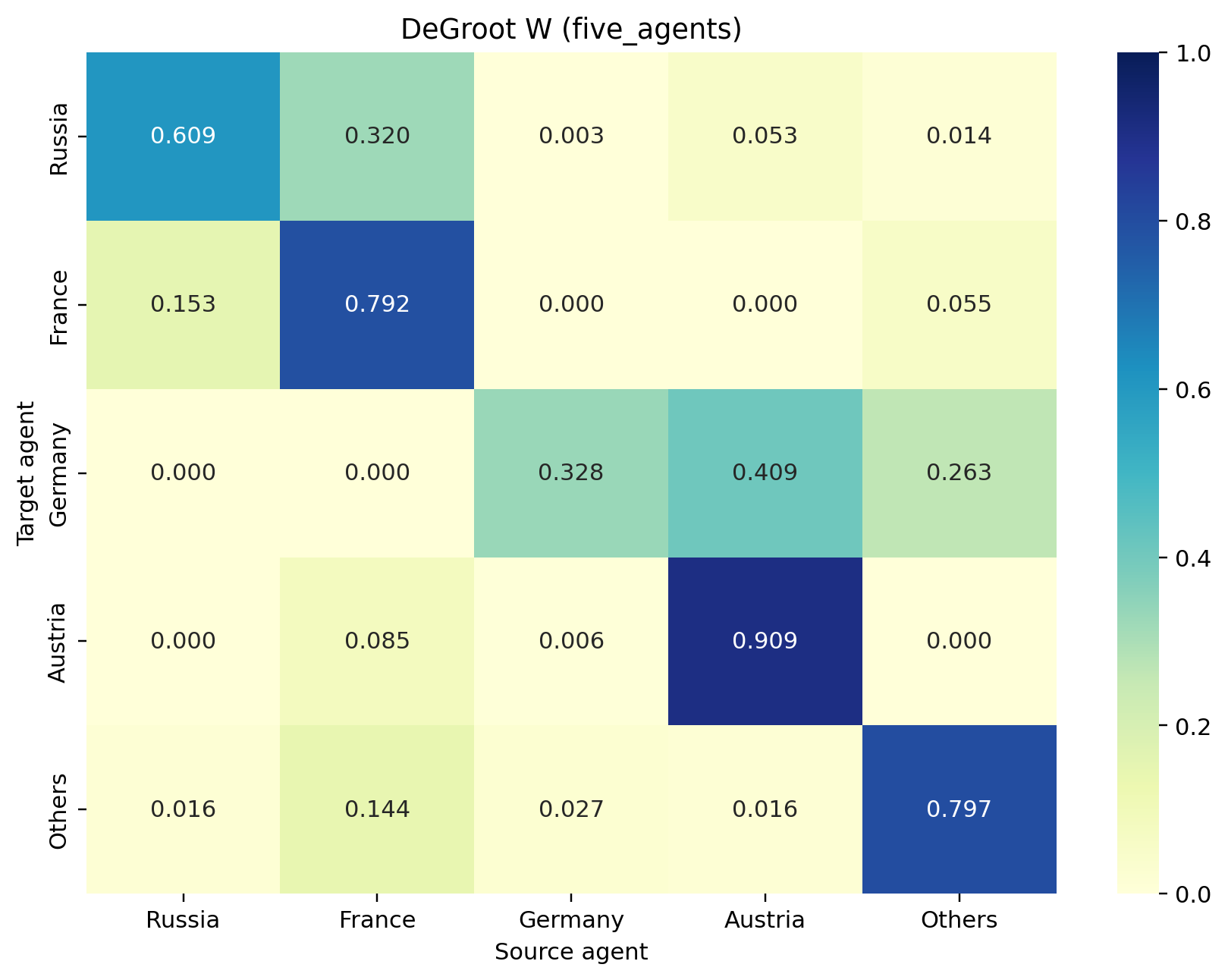}
    \caption{Five-agent DG matrix}
    \label{fig:midi_five_agent_DG}
\end{subfigure}
\hfill
\begin{subfigure}{0.47\textwidth}
    \centering
    \includegraphics[width=\textwidth]{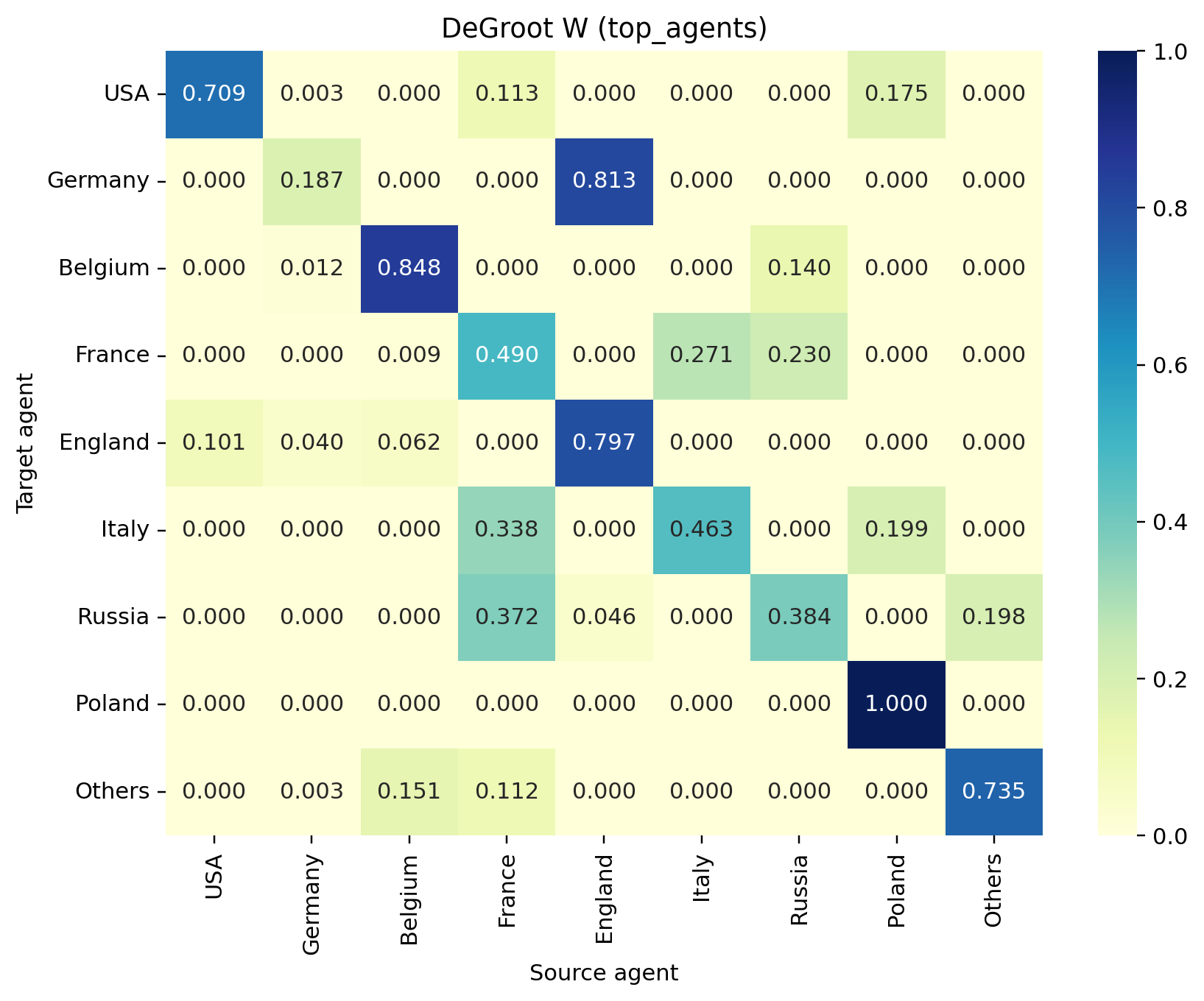}
    \caption{Top-agent DG matrix}
    \label{fig:midi_top_agent_DG}
\end{subfigure}
\caption{Estimated DG influence matrices in the MIDI BiLSTM stress test.
Rows denote target agents and columns denote source agents. Because the
underlying opinion space is dominated by the Modern class, these matrices are
reported as diagnostic outputs rather than as historical cross-national
influence estimates.}
\label{fig:midi_DG_matrices}
\end{figure}

For completeness, Figure~\ref{fig:midi_DG_matrices} reports the estimated
DG influence matrices for both the five-agent and top-agent designs. These
matrices are included to show that the estimation procedure remains numerically
well-defined under the MIDI BiLSTM representation. However, because the
underlying opinion space is dominated by the Modern class and the forecasts do
not improve over persistence, we do not interpret these matrices as historical
evidence of cross-national musical influence.

The MIDI stress test has two implications. First, the proposed
state-construction and dynamics-estimation procedure can be applied to a
separate symbolic-MIDI corpus and to a different representation space. Second,
supervised era-classifier probabilities are not a suitable replacement for the
ChordBERT/PCA harmonic representation used in the main analysis. Because the
BiLSTM opinion vectors collapse toward one dominant class and fail to improve
over persistence, we treat the MIDI experiment as a diagnostic stress test rather
than as evidence for the main historical influence estimates.

\end{document}